\documentclass[12 pt]{article}

\pdfoutput=1

\usepackage{fullpage}
\usepackage{amscd}
\usepackage{amsmath}
\usepackage{amssymb}
\usepackage{amsthm}
\usepackage{graphicx}
\usepackage{latexsym}
\usepackage{verbatim}
\input epsf.tex

\newtheorem{theorem}{Theorem}[section]
\newtheorem{lemma}[theorem]{Lemma}
\newtheorem{corollary}[theorem]{Corollary}

\newtheorem{proposition}[theorem]{Proposition}

\theoremstyle{definition}

\theoremstyle{definition}
\newtheorem{algorithm}[theorem]{Algorithm}

\theoremstyle{remark}
\newtheorem{example}[theorem]{Example}

\theoremstyle{definition} 
\newtheorem{definition}[theorem]{Definition}

\newcounter{tempthm}
\newcounter{tempsec}

\newcommand{\savecounter}[1]{\newcounter{thmcounter#1}
\setcounter{thmcounter#1}{\value{theorem}}
\newcounter{seccounter#1}
\setcounter{seccounter#1}{\value{section}}}

\newcommand{\usesavedcounter}[1]{\setcounter{tempthm}{\value{theorem}}
\setcounter{theorem}{\value{thmcounter#1}}
\setcounter{tempsec}{\value{section}}
\setcounter{section}{\value{seccounter#1}}}

\newcommand{\restorecounter}{\setcounter{theorem}{\value{tempthm}}
\setcounter{section}{\value{tempsec}}}

\newcommand{\Nm}{\mathbb{N}}

\newcommand{\Rm}{\mathbb{R}}

\newcommand{\shortv}[1]{}

\DeclareMathOperator{\prob}{Prob}

\title{\LARGE \bf
Correlated Equilibria in Continuous Games: Characterization and Computation
}

\author{Noah D. Stein, Pablo A. Parrilo, and Asuman Ozdaglar
\thanks{Department of Electrical Engineering,
        Massachusetts Institute of Technology: Cambridge, MA 02139.
        {\tt\small nstein@mit.edu}, {\tt\small parrilo@mit.edu}, and {\tt\small asuman@mit.edu}.}
\thanks{This research was funded in part by National Science Foundation grants $\text{DMI-}0545910$ and ECCS-$0621922$ and AFOSR MURI subaward $2003\text{-}07688\text{-}1$.}
      }

\begin{document}

\markright{LIDS Technical Report $2805$}

\maketitle

\thispagestyle{headings}

\pagestyle{plain}

\begin{abstract}
We present several new characterizations of correlated equilibria in games with continuous utility functions.  These have the advantage of being more computationally and analytically tractable than the standard definition in terms of departure functions.  We use these characterizations to construct effective algorithms for approximating a single correlated equilibrium or the entire set of correlated equilibria of a game with polynomial utility functions.
\end{abstract}


\section{Introduction}

In finite games correlated equilibria are simpler than Nash equilibria in several senses -- mathematically at least, if not conceptually.  The set of correlated equilibria is a convex polytope, described by finitely many explicit linear inequalities, while the set of (mixed) Nash equilibria need not be convex or connected and can contain components which look like essentially any real algebraic variety (set described by polynomial equations on real variables) \cite{d:une}.  The existence of correlated equilibria can be proven by elementary means (linear programming or game theoretic duality \cite{hs:ece}), whereas the existence of Nash equilibria seems to require nonconstructive methods (e.g., fixed point theorems as in \cite{nash:ncg,glicksberg:cg}) or the analysis of complicated algorithms \cite{lh:epbg,fy:rt,gl:gncfyrt}.  Computing a sample correlated equilibrium or a correlated equilibrium optimizing some quantity such as social welfare can be done efficiently \cite{gz:ncescc,p:ccempg}; strong evidence in complexity theory suggests that the corresponding problems for Nash equilibria are hard \cite{dgp:ccne,cd:nashcomp, gz:ncescc}.

There are several exceptional classes of games for which the above problems about Nash equilibria become easy.  The most important here are the zero-sum games.  Broadly speaking, Nash equilibria of these games have complexity similar to correlated equilibria of general games.   In particular, the set of Nash equilibria is an easily described convex polytope, existence can be proven by duality, and a sample equilibrium can be computed efficiently.

The situation in games with infinite strategy sets is not nearly so clear.  For the computational sections of this paper we restrict attention to the simplest such class of games, those with finitely many players, strategy sets equal to $[-1,1]$, and polynomial utility functions.  We make this restriction for several reasons.

The first is conceptual and notational simplicity.  Results similar to ours will hold when the strategy sets are general compact semialgebraic (described by finitely many polynomial equations and inequalities) subsets of $\Rm^n$.  However, dealing with this additional level of generality requires machinery from computational real algebraic geometry and does little to illuminate our basic methods.

The second is generality.  Much of the study of games with infinite strategy sets is fraught with assumptions of concavity or quasiconcavity which appear to be motivated not by natural game theoretic premises, but rather by the inadequacy of available tools for games without these properties.  While polynomiality assumptions and concavity assumptions are both rigid in their own ways, polynomials have the benefit of being dense in the space of all continuous functions, and thus suitable for approximating a much wider class of games.

The third reason is convenience.  The algebraic structure we gain by restricting attention to polynomials allows us to use recent advances in semidefinite programming and real algebraic geometry to construct efficient algorithms and gain conceptual insights.

Little is known about correlated equilibria of these polynomial games, but much is known about Nash equilibria.  Most importantly, the set of mixed Nash equilibria is nonempty and admits a finite-dimensional description in terms of the moments of the players' mixed strategies \cite{sop:slrcg}.

This set of moments can be described explicitly in terms of polynomial equations and inequalities \cite{sop:slrcg}.  The Nash equilibrium conditions are expressible via first order statements, so the set of all moments of Nash equilibria is a real algebraic variety and can be computed in theory, albeit not efficiently in general.  In the two-player zero-sum case, the set of Nash equilibria can be described by a semidefinite program (an SDP is a generalization of a linear program which can be efficiently solved; see the appendix), hence we can compute a sample Nash equilibrium or one which optimizes some linear functional in polynomial time \cite{pp:polygames}.  A summary of the results described so far is shown in Table \ref{tab:eqsummary}.

\begin{table}
\label{tab:eqsummary}
\begin{center}
\begin{tabular}{c|c|c|c|}
 & Nash equilibria & Nash equilibria & correlated equilibria \\
 & (non-zero sum) & (zero sum) & \\
 \hline
 Finite games & Semialgebraic set \cite{lm:nevpe} & LP & LP \cite{a:ceebr}\\
 \hline
 Polynomial games & Semialgebraic set \cite{sop:slrcg} & SDP \cite{pp:polygames} & ? \\
 \hline
\end{tabular}
\end{center}
\caption{Comparison of the simplest known description of different classes of equilibrium sets in finite and polynomial games.}
\end{table}

\paragraph{Contributions}The impetus for this paper was to address the bottom right cell of Table \ref{tab:eqsummary}, the one with the question mark.  The table seems to suggest that the set of correlated equilibria of a polynomial game should be describable by a semidefinite program.  We will see that this is approximately true, but not exactly.  The contribution of this paper is twofold.

\begin{itemize}
\item First, we present several new characterizations of correlated equilibria in games with continuous utility functions (polynomiality is not needed here).  In particular we show that the standard definition of correlated equilibria in terms of measurable departure functions is equivalent to other definitions in which the utilities are integrated against all test functions in some class (Theorem \ref{thm:correqmainchar}).  This characterization does not have any obvious game theoretic significance, but it is extremely useful analytically and it forms the base for our other contributions.

\item Second, we present several algorithms for approximating correlated equilibria within an arbitrary degree of accuracy.  We present one inefficient linear programming based method as a benchmark, followed by two semidefinite programming based algorithms which perform much better in practice.  The first SDP algorithm, called adaptive discretization, iteratively computes a sequence of approximate correlated equilibria supported on finite sets (Section \ref{sec:adaptivedisc}).  We enlarge the support sets at each iteration using a heuristic which guarantees convergence in general and yields fast convergence in practice.  The second SDP algorithm, called moment relaxation, does not discretize the strategy spaces but instead works in terms of joint moments.  It produces a nested sequence of outer approximations to the set of joint moments of correlated equilibrium distributions, and these approximate equilibrium sets are described by semidefinite programs (Section \ref{subsec:moment}).  These relaxations depend crucially on one of the correlated equilibrium characterizations we have developed.
\end{itemize}

\paragraph{Related literature} The questions we address and the techniques we use are inspired by existing literature in two main areas.  First, our work is related to a number of papers in the game theory literature.
\begin{itemize}
\item Aumann defines correlated equilibria in his famous paper \cite{a:scrs}, focusing on finite games to establish the basic properties and important examples.  He obtains existence as a consequence of Nash's theorem on the existence of Nash equilibria in finite games \cite{nash:ncg}.

\item Hart and Schmeidler show that existence of correlated equilibria in finite games can be proven directly by a duality argument \cite{hs:ece}.  They then use a careful limiting argument to prove existence of correlated equilibria in continuous games with compact Hausdorff strategy spaces (Theorem $3$ of that paper).  The germs of ideas in this limiting argument are developed further in Section \ref{sec:char} of the present paper to yield various characterizations of correlated equilibria.  It is worth noting that in \cite{hs:ece} the authors also consider part \eqref{item:char} of Corollary \ref{cor:correqmultiplierchar} as a candidate definition of correlated equilibria.  They discard it is as not obviously capturing the game theoretic idea of correlated equilibrium, but we prove that it is nonetheless an equivalent definition in games with continuous utilities.

\item Stoltz and Lugosi study learning algorithms which converge to correlated equilibria in continuous games \cite{sl:lcegcss}.  These algorithms have a game theoretic interpretation as avoiding ``regret'' in a repeated game setting.  Each player can carry out these procedures separately without knowledge of his opponents' utilities.  These are conceptual advantages over our methods, which merely aim for efficient computation.

However, these advantages come at a cost.  The learning procedures require each player to solve a fixed point equation at each iteration.  In general finding fixed points is as hard as finding Nash equilibria \cite{dgp:ccne,cd:nashcomp}, so these procedures do not seem to lead directly to efficient methods for computing correlated equilibria of continuous games.

There exist classes of fixed point equations which can be solved efficiently (i.e., the steady-state distribution of a Markov chain, which is defined by a linear program).  To our knowledge there has been no work on whether the equations of \cite{sl:lcegcss} fall into such a class.

Furthermore, each of these learning algorithms either makes concavity-type assumptions about the utility functions, which we seek to avoid for modeling flexibility, or discretizes the players' strategy spaces a priori.  We will see in Section \ref{sec:staticdisc} that such discretization without regard to the structure of the game can result in slow convergence.

However, we will make use of some of the tools which Stoltz and Lugosi have created.  In particular, they consider replacing the class of all measurable departure functions with a smaller class, such as simple or continuous departure functions, and study when this yields an equivalent equilibrium notion.  One result of this type is stated as Lemma \ref{lem:simplecorreq} below and used to prove our characterization theorems.

\item Germano and Lugosi prove the existence of correlated equilibria with small support in finite games \cite{gl:essce}.  To prove this they analyze the extreme points of the set of correlated equilibria.  Such an analysis cannot carry over directly to polynomial games because the set of correlated equilibria of polynomial games may have extreme points with arbitrarily large finite support or with infinite support \cite{sop:sece}.  Support bounds for correlated equilibria of polynomial games are proven in \cite{s:mastersthesis} using similar tools, but assuming a finitely supported Nash equilibrium is on hand as a starting point.  Since Nash equilibria are generally assumed to be harder to compute than correlated equilibria, these results do not apply in the present setting where the goal is efficient computation.

\item Separately from the literature on correlated equilibria, Dresher, Karlin, and Shapley study the structure of Nash equilibria in zero-sum games with polynomial or separable (polynomial-like) utility functions.  They show how to cast separable games as finite-dimensional ``convex games'' by replacing the infinite-dimensional mixed strategy spaces with finite-dimensional spaces of moments \cite{dks:pg} and prove existence of equilibria via fixed point arguments \cite{dk:scgfp}.  There always exist finitely supported equilibria in separable games as can be shown using the finite-dimensonality of the moment spaces.  The rich geometry of these spaces is studied in \cite{ks:gms}.  Most of these results as well as ad hoc methods for computing equilibria in simple cases are summarized in Karlin's book \cite{karlin:tig}.  The authors of the present paper study generalizations and extensions of these results to nonzero-sum separable games in \cite{sop:slrcg}.
\end{itemize}

Second, our work is related to results from the optimization and computer science literature.

\begin{itemize}
\item Aumann showed that the set of correlated equilibria of a finite game is defined by polynomially many (in the size of the payoff tables) linear inequalities \cite{a:ceebr}.  However, it was not clear whether this meant they could be computed in polynomial time.  This question was settled in the affirmative when Khachian proved that linear programs could be solved in polynomial time; for an overview of this and other more efficient algorithms, see \cite{bt:ilo}.  Papadimitriou extends this result in \cite{p:ccempg}, showing that correlated equilibria can be computed efficiently in many classes of games for which the payoffs can be written succinctly, even if the explicit payoff tables would be exponential in size.

\item The breakthrough in optimization most directly related to the work in this paper is the development of semidefinite programming, a far-reaching generalization of linear programming which is still polynomial-time solvable (for an overview, see the appendix and \cite{vb:sdp}).  More specifically, the development of sum of squares methods has allowed many optimization problems involving polynomials or moments of measures to be solved efficiently \cite{p:phd}.  Parrilo applies these techniques to efficiently compute Nash equilibria of two-player zero-sum polynomial games in \cite{pp:polygames}.
\end{itemize}

The remainder of this paper is organized as follows.  In Section \ref{sec:char} we define the classes of games we study and correlated equilibria thereof, then prove several characterization theorems.  We present algorithms for approximating sample correlated equilibria and the set of correlated equilibria of polynomial games in Section \ref{sec:comp}.  Finally, we close with conclusions and directions for future work.

\section{Characterizations of Correlated Equilibria}
\label{sec:char}
In this section we will define finite and continuous games along with correlated equilibria thereof.  We will present several known characterizations of correlated equilibria in finite games and show how these naturally extend to continuous games.

Some notational conventions used throughout are that subscripts refer to players, while superscripts are frequently used for other indices (it will be clear from the context when they represent exponents).  If $S_j$ are sets for $j=1,\ldots,n$ then $S = \Pi_{j=1}^n S_j$ and $S_{-i} = \Pi_{j\neq i} S_j$.  The $n$-tuple $s$ and the $(n-1)$-tuple $s_{-i}$ are formed from the points $s_j$ similarly.  The set of regular Borel probability measures $\pi$ over a compact Hausdorff space $S$ is denoted by $\Delta(S)$.  For simplicity we will write $\pi(s)$ in place of $\pi(\{s\})$ for the measure of a singleton $\{s\}\subseteq S$.  All polynomials will be assumed to have real coefficients.

\subsection{Finite Games}
\label{subsec:finite}
We start with the definition of a finite game.

\begin{definition}
A \textbf{finite game} consists of \textbf{players} $i = 1,\ldots, n$, each of whom has a finite \textbf{pure strategy set} $C_i$ and a \textbf{utility} or \textbf{payoff function} $u_i: C\rightarrow \Rm$, where $C = \Pi_{j=1}^n C_j$.
\end{definition}

Each player's objective is to maximize his (expected) utility.  We now consider what it would mean for the players to maximize their utility if their strategy choices were correlated.  Let $R$ be a random variable taking values in $C$ distributed according to some measure $\pi\in\Delta(C)$.  A realization of $R$ is a \textbf{pure strategy profile} (a choice of pure strategy for each player) and the $i^{\text{th}}$ component of the realization $R_i$ will be called the recommendation to player $i$.  Given such a recommendation, player $i$ can use conditional probability to form a posteriori beliefs about the recommendations given to the other players.  A distribution $\pi$ is defined to be a correlated equilibrium if no player can ever expect to unilaterally gain by deviating from his recommendation, assuming the other players play according to their recommendations.

\begin{definition}
\label{def:finitegamecorreq}A \textbf{correlated equilibrium} of a finite game is a joint probability measure $\pi\in\Delta(C)$ such that if $R$ is a random variable distributed according to $\pi$ then
\begin{equation*}
\label{eq:conditionalcorreq}
\mathbb{E}\left[u_i(t_i,R_{-i}) - u_i(R)\vert R_i = s_i\right] \equiv \sum_{s_{-i}\in C_{-i}} \prob(R = s | R_i = s_i)\left[u_i(t_i,s_{-i})-u_i(s)\right] \leq 0
\end{equation*}
for all players $i$, all $s_i\in C_i$ such that $\prob(R_i = s_i) > 0$, and all $t_i\in C_i$.  
\end{definition}

While this definition captures the idea we have described above, the following characterization is easier to apply and visualize.
\begin{proposition}
\label{prop:finitecorreqchar}
A joint probability measure $\pi\in\Delta(C)$ is a correlated equilibrium of a finite game if and only if
\begin{equation}
\label{eq:finitecorreqcond}
\sum_{s_{-i}\in C_{-i}} \pi(s) \left[u_i(t_i,s_{-i}) - u_i(s)\right] \leq 0
\end{equation}
for all players $i$ and all $s_i,t_i\in C_i$.
\end{proposition}


This proposition shows that the set of correlated equilibria is defined by a finite number of linear equations and inequalities (those in \eqref{eq:finitecorreqcond} along with $\pi(s) \geq 0$ for all $s\in C$ and $\sum_{s\in C} \pi(s) = 1$) and is therefore convex and even polyhedral.  It can be shown via linear programming duality that this set is nonempty \cite{hs:ece}.  This can be shown alternatively by appealing to the fact that Nash equilibria exist and are the same as correlated equilibria which are product distributions.

We can think of correlated equilibria as joint distributions corresponding to recommendations which will be given to the players as part of an extended game.  The players are then free to play any function of their recommendation as their strategy in the game.
\begin{definition}
A function $\zeta_i: C_i\rightarrow C_i$ is called a \textbf{departure function}.
\end{definition}
If it is a Nash equilibrium of this extended game for each player to play his recommended strategy (i.e. if no player has an incentive to unilaterally deviate from using the identity departure function), then the distribution is a correlated equilibrium.  This interpretation is due to Aumann \cite{a:ceebr} and is justified by the following alternative characterization of correlated equilibria.

\begin{proposition}
\label{prop:finitecorreqchar2}
A joint probability measure $\pi\in\Delta(C)$ is a correlated equilibrium of a finite game if and only if
\begin{equation}
\label{eq:finitecorreqcond2}
\sum_{s\in C} \pi(s)\left[u_i(\zeta_i(s_i),s_{-i})-u_i(s)\right] \leq 0
\end{equation}
for all players $i$ and all departure functions $\zeta_i$.
\end{proposition}

For examples and more discussion of the basics of correlated equilibria, including the ideas behind the equivalence of these characterizations, see \cite{a:scrs,a:ceebr}.


\subsection{Continuous Games}
Again we begin with the definition of this class of games.
\begin{definition}
A \textbf{continuous game} consists of an arbitrary (possibly infinite) set $I$ of players $i$, each of whom has a pure strategy set $C_i$ which is a compact Hausdorff space and a utility function $u_i: C\rightarrow \Rm$ which is continuous.
\end{definition}

Note that any finite set forms a compact Hausdorff space under the discrete topology and any function out of such a set is continuous, so the class of continuous games includes the finite games.  Another class of continuous games are the polynomial games, which are our primary focus when we study computation of correlated equilibria in the sections which follow.  The theorems and proofs below can safely be read with polynomial games in mind, ignoring such topological subtleties as regularity of measures.  However the extra generality of arbitrary continuous games requires little additional work in the proofs of the characterization theorems, so we will not formally restrict our attention to polynomial games here.

\begin{definition}
A \textbf{polynomial game} is a continuous game with $n < \infty$ players in which the pure strategy spaces are $C_i = [-1,1]$ for all players and the utility functions are polynomials.
\end{definition}

Defining correlated equilibria in continuous games requires somewhat more care than in finite games.  Because of the technical difficulties of dealing with conditional distributions on continuous spaces, it is preferable not to formulate our new definition by generalizing Definition \ref{def:finitegamecorreq} directly.  An obvious thing to try would be to replace the sum in Proposition \ref{prop:finitecorreqchar} with an integral and to choose that as the definition of correlated equilibria in continuous games.  That would be simple enough, but this leads to a notion which is very weak and uninformative.  Since we would be integrating over ``slices'' our candidate definition would be met, for example, by any continuous probability distribution regardless of the game chosen.  Thus we have to use a different approach.

The standard definition of correlated equilibria in continuous games (as used in \cite{hs:ece}) instead follows Proposition \ref{prop:finitecorreqchar2}.  In this case we must add the additional assumption that the departure functions be Borel\footnote{The Borel $\sigma$-algebra is the $\sigma$-algebra generated by the topology on $C_i$, which was assumed given in the definition of a continuous game.} measurable to ensure that the integrals are defined.  For finite games this assumption is vacuous so this definition is equivalent to Definition \ref{def:finitegamecorreq}.
\begin{definition} 
\label{def:correq}
A \textbf{correlated equilibrium} of a continuous game is a joint probability measure $\pi\in\Delta(C)$ such that
\begin{equation*}
\int\left[u_i(\zeta_i(s_i),s_{-i}) - u_i(s)\right]\,d\pi(s) \leq 0
\end{equation*}
for all $i$ and all Borel measurable departure functions $\zeta_i$.
\end{definition}

Before stating and proving alternative characterization theorems, we will discuss some of the difficulties of working with correlated equilibria in continuous games and this definition in particular.  The goal here is to motivate the need for alternative characterizations.

The problem of computing Nash equilibria of polynomial games can be formulated exactly as a finite-dimensional nonlinear program or as a system of polynomial equations and inequalities \cite{sop:slrcg}.  The key feature of the problem which makes this possible is the fact that it has an explicit finite-dimensional formulation in terms of the moments of the players' mixed strategies.

To see this, suppose that player $1$ chooses his action $x\in [-1,1]$ according to a mixed strategy $\sigma$ (a probability distribution over $[-1,1]$).  Each player's utility function is a multivariate polynomial which only contains terms whose degree in $x$ is at most some constant integer $d$.  Then regardless of how everyone chooses their strategies, their expected utility will only depend on $\sigma$ through the moments $\int x\,d\sigma(x),\int x^2\,d\sigma(x),\ldots, \int x^d\,d\sigma(x)$.  Therefore player $1$ can switch from $\sigma$ to any other mixed strategy with the same first $d$ moments without affecting game play, and we can think of the Nash equilibrium problem as one in which each player seeks to choose moments which correspond to an actual probability distribution and form a Nash equilibrium.

On the other hand there is no exact finite-dimensional characterization of the set of correlated equilibria in polynomial games; for a counterexample see \cite{sop:sece}.  Given the characterization of Nash equilibria in terms of moments, a natural attempt would be to try to characterize correlated equilibria in terms of the joint moments, i.e. the values $\int s_1^{k_1}\cdots s_n^{k_n}\,d\pi$ for nonnegative integers $k_i$ and joint measures $\pi$.  In fact we will be able to obtain such a characterization below, albeit in terms of infinitely many joint moments.  The reason this attempt fails to yield a finite dimensional formulation is that the definition of a correlated equilibrium implicitly imposes constraints on the conditional distributions of the equilibrium measure.  A finite set of moments does not contain enough information about these conditional distributions to check the required constraints exactly.  Therefore we also consider approximate correlated equilibria.

\begin{definition}
\label{def:epscorreq}
An \textbf{$\epsilon$-correlated equilibrium} of a continuous game is a joint probability measure $\pi\in\Delta(C)$ such that
\begin{equation*}
\int\left[u_i(\zeta_i(s_i),s_{-i}) - u_i(s)\right]\,d\pi(s) \leq \epsilon
\end{equation*}
for all $i$ and all Borel measurable departure functions $\zeta_i$.  This definition reduces to that of a correlated equilibrium when $\epsilon = 0$.
\end{definition}


That is to say, $\epsilon$-correlated equilibria are distributions of recommendations in which no player can improve his expected payoff by more than $\epsilon$ by deviating from his recommendation unilaterally.  Compare this definition to the main characterization theorem for $\epsilon$-correlated equilibria below (Theorem \ref{thm:correqmainchar}).  This theorem shows that $\epsilon$-correlated equilibria can equivalently be defined by integrating the utilities against any sufficiently rich class of test functions, instead of by using measurable departure functions.  Intuitively, the advantage of this characterization is that the product $f_i(s_i)u_i(t_i,s_{-i})$ is a ``simpler'' mathematical object than the composition $u_i(\zeta_i(s_i),s_{-i})$, especially when $f_i$, $t_i$, and $\zeta_i$ are allowed to vary.  While this characterization does not have an obvious game theoretic interpretation, it allows us to compute correlated equilibria both algorithmically (Section \ref{sec:comp}) and analytically \cite{sop:sece}.

There also exist a variety of characterizations in which the departure functions are restricted to lie in a particular class (e.g., Lemma \ref{lem:simplecorreq} below and similar results in \cite{sl:lcegcss}) and no test functions are used.  These characterizations have the advantages of conceptual simplicity and ease of interpretation.  However, any characterization involving departure functions suffers from the difficulty that compositions of the utilities and the departure functions must be computed and these will likely be complex even if the departure functions are restricted to a simple class.  The difficulty is magnified by the fact that even these restricted classes of departure functions are large and often difficult to parametrize in a way which is amenable to computation.  Therefore it seems that departure function characterizations of correlated equilibria cannot be applied directly to yield effective computational procedures.

\begin{theorem}
\label{thm:correqmainchar}
A probability measure $\pi\in\Delta(C)$ is an $\epsilon$-correlated equilibrium of a continuous game if and only if for all players $i$, positive integers $k$, strategies $t_i^1,\ldots,t_i^k\in C_i$, and functions $f_i^1,\ldots,f_i^k: C_i\rightarrow [0,1]$ in one of the classes
\begin{enumerate}
\item \label{item:char} Weighted measurable characteristic functions,
\item \label{item:simp} Measurable simple functions (i.e., functions with finite range),
\item \label{item:meas} Measurable functions,
\item \label{item:cont} Continuous functions,
\item \label{item:poly} Squares of polynomials (if $C_i\subset \Rm^{k_i}$ for some $k_i$ for all $i$).
\end{enumerate}
such that $\sum_{j=1}^k f_i^j(s_i)\leq 1$ for all $s_i\in C_i$, the inequality
\begin{equation}
\label{eq:correqmainchar}
\sum_{j=1}^k\int f_i^j(s_i)\left[u_i(t_i^j,s_{-i})-u_i(s)\right]\,d\pi \leq \epsilon
\end{equation}
holds.
\end{theorem}

To prove this, we need several approximation lemmas.

\begin{lemma}[A special case of Lemma $20$ in \cite{sl:lcegcss}]
\label{lem:simplecorreq}
Simple departure functions (those with finite range) suffice to define $\epsilon$-correlated equilibria in continuous games.  That is to say, a joint measure $\pi$ is an $\epsilon$-correlated equilibrium if and only if
\begin{equation*}
\int\left[u_i(\xi_i(s_i),s_{-i}) - u_i(s)\right]\,d\pi(s) \leq \epsilon
\end{equation*}
for all players $i$ and all Borel measurable simple departure functions $\xi_i$.
\end{lemma}

\begin{proof}
The forward direction is trivial.  To prove the reverse, first fix $i$.  Then choose any measurable departure function $\zeta_i$ and let $\delta>0$ be arbitrary.  By the continuity of $u_i$ and compactness of the strategy spaces there exists a finite open cover $U^1,\ldots,U^k$ of $C_i$ such that $s_i,s'_i\in U^j$ implies $\lvert u_i(s_i,s_{-i}) - u_i(s'_i,s_{-i})\rvert < \delta$ for all $s_{-i}\in C_{-i}$ and $j=1,\ldots, k$.  Fix any $s_i^j\in U^j$ for all $j$.  Define a simple measurable departure function $\xi_i$ by $\xi_i(s_i) = s_i^j$ where $j = \min \{l: \zeta_i(s_i)\in U^l\}$.  Then $\lvert u_i(\zeta_i(s_i),s_{-i})-u_i(\xi_i(s_i),s_{-i})\rvert<\delta$ for all $s\in C$, so
\begin{equation*}
\begin{split}
\int & \left[u_i(\zeta_i(s_i), s_{-i})-u_i(s)\right]\,d\pi(s)
 \leq \int\left[u_i(\xi_i(s_i),s_{-i}) + \delta -u_i(s)\right]\,d\pi(s)
 \leq \epsilon + \delta.
\end{split}
\end{equation*}
Letting $\delta$ go to zero completes the proof.
\end{proof}
%


\begin{lemma}
\label{lem:generallusin}
If $C$ is a compact Hausdorff space, $\mu$ is a finite regular Borel measure on $C$, $f^1,\ldots,f^k: C\rightarrow [0,1]$ are measurable functions such that $\sum_{j=1}^k f^j \leq 1$, and $\delta > 0$, then there exist continuous functions $g^1,\ldots,g^k: C\rightarrow [0,1]$ such that $\mu(\{x\in C: f^j(x)\neq g^j(x)\}) < \delta$ for all $j$ and $\sum_{j=1}^k g^j \leq 1$.
\end{lemma}

\begin{proof}
We can apply Lusin's theorem which states exactly this result in the case $k = 1$ \cite{r:rca}.  If $k > 1$, then we can apply the $k=1$ case with $\frac{\delta}{k}$ in place of $\delta$ to each of the $f^j$.  Call the resulting continuous functions $\tilde{g}^j$.  Then $\mu(\{x\in C:f^j(x)\neq \tilde{g}^j(x)\text{ for some }j\}) < \delta$.  But $\sum_{j=1}^k f^j \leq 1$, so $\mu(\{x\in C: \sum_{j=1}^k \tilde{g}^j(x) > 1\}) <\delta$.  Let $h(x) = \max\{1,\sum_{j=1}^k \tilde{g}^j(x)\}$ so $h: C\rightarrow [1,\infty)$ is a continuous map.  Define $g^j(x) = \frac{\tilde{g}^j(x)}{h(x)}$.  Then the $g^j$ are continuous, sum to at most unity, and are equal to the $f^j$ wherever all of the $\tilde{g}^j$ equal the $f^j$, i.e. except on a set of measure at most $\delta$.
\end{proof}

\begin{lemma}
\label{lem:contpolyapprox}
If $C\subset\Rm^d$ is compact, $f^1,\ldots,f^k: C\rightarrow [0,1]$ are continuous functions such that $\sum_{j=1}^k f^j \leq 1$, and $\delta > 0$, then there exist polynomials $p^1,\ldots,p^k: C\rightarrow [0,1]$ which are squares such that $\lvert f^j(x) - p^j(x)\rvert \leq \delta$ for all $x\in C$ and $\sum_{j=1}^k p^j \leq 1$.
\end{lemma}

\begin{proof}
By the Stone-Weierstrass theorem, any continuous function on a compact subset of $\Rm^d$ can be approximated by a polynomial arbitrarily well with respect to the sup norm.  Approximating the square root of a nonnegative function $f$ using this theorem and squaring the resulting polynomial shows that a nonnegative continuous function on a compact subset of $\Rm^d$ can be approximated arbitrarily well by a square of a polynomial with respect to the sup norm.

Let $\tilde{p}^j$ be a square of a polynomial which approximates $f^j$ within $\frac{\delta}{2k}$ in the sup norm.  Since $f^j$ takes values in $[0,1]$, $\tilde{p}^j$ takes values in $\left[0,1+\frac{\delta}{2k}\right]$.  Let $p^j = \frac{\tilde{p}^j}{1+\frac{\delta}{2}}$.  Then for all $x\in C$ we have $p^j(x)\leq \tilde{p}^j(x)$ and
\begin{equation*}
\tilde{p}^j(x) - p^j(x) = \tilde{p}^j(x) - \frac{\tilde{p}^j(x)}{1+\frac{\delta}{2}} = \tilde{p}^j(x)\left(\frac{\frac{\delta}{2}}{1+\frac{\delta}{2}}\right) \leq \left(1+\frac{\delta}{2k}\right)\left(\frac{\frac{\delta}{2}}{1+\frac{\delta}{2}}\right) \leq \frac{\delta}{2},
\end{equation*}
so $p^j(x)$ is within $\frac{\delta}{2}$ of $\tilde{p}^j(x)$ for all $x\in C$.  By the triangle inequality $p^j$ approximates $f^j$ within $\delta$ in the sup norm.  Furthermore for all $x\in C$ we have
\begin{equation*}
\sum_{j=1}^k p^j(x) = \frac{1}{1+\frac{\delta}{2}}\sum_{j=1}^k \tilde{p}^j(x) \leq \frac{1}{1+\frac{\delta}{2}}\sum_{j=1}^k \left(f^j(x)+\frac{\delta}{2k}\right)\leq \frac{1}{1+\frac{\delta}{2}}\left(1+\frac{\delta}{2}\right) = 1.\qedhere
\end{equation*}
\end{proof}

\begin{proof}[Proof of Theorem \ref{thm:correqmainchar}]
First we prove that if $\pi$ is an $\epsilon$-correlated equilibrium then \eqref{eq:correqmainchar} holds in the case where the $f_i^j$ are simple.  We can choose a partition $B_i^1,\ldots,B_i^l$ of $C_i$ into disjoint measurable sets such that $f_i^j = \sum_{m=1}^l c_{jm}\chi_{B_i^m}$ where $c_{jm}\in [0,1]$ and $\chi_{B_i^m}$ denotes the indicator function which is unity on $B_i^m$ and zero elsewhere.  Define a departure function $\zeta_i:C_i\rightarrow C_i$ piecewise on the $B_i^m$ as follows.  If
\begin{equation*}
\int_{B_i^m\times C_{-i}}\left[u_i(t_i^j,s_{-i})-u_i(s)\right]\,d\pi
\end{equation*}
is nonnegative for some $j$ define $\zeta_i(s_i) = t_i^j$ for all $s_i\in B_i^m$ where $j$ is chosen to maximize the above integral.  If the integral is negative for all $j$ define $\zeta_i(s_i) = s_i$ for all $s_i\in B_i^m$.  Then we have
\begin{equation*}
\sum_{j=1}^k c_{jm}\int_{B_i^m\times C_{-i}}\left[u_i(t_i^j,s_{-i})-u_i(s)\right]\,d\pi\leq \int_{B_i^m\times C_{-i}}\left[u_i(\zeta_i(s_i),s_{-i})-u_i(s)\right]\,d\pi
\end{equation*}
for all $m$.  Summing over $m$ and using the definition of an $\epsilon$-correlated equilibrium yields \eqref{eq:correqmainchar} in the case where the $f_i^j$ are simple.

Conversely suppose that \eqref{eq:correqmainchar} holds for all measurable simple functions.  Let $\zeta_i: C_i\rightarrow C_i$ be any simple departure function.  Let $t_i^1,\ldots,t_i^k$ be the range of $\zeta_i$ and $B_i^j = \zeta_i^{-1}(\{t_i^j\})$.  Defining $f_i^j = \chi_{B_i^j}$, \eqref{eq:correqmainchar} says exactly that $\pi$ satisfies the $\epsilon$-correlated equilibrium condition for the departure function $\zeta_i$.  By Lemma \ref{lem:simplecorreq}, $\pi$ is an $\epsilon$-correlated equilibrium.

Any simple function can be written as a sum of weighted characteristic functions, so by making several of the $t_i^j$ the same, we see that \eqref{eq:correqmainchar} for weighted characteristic functions is the same as \eqref{eq:correqmainchar} for simple measurable functions.  If the inequality \eqref{eq:correqmainchar} holds for all simple measurable functions, a standard limiting argument proves that it holds for all measurable $f_i^j$, hence for all continuous $f_i^j$.

Suppose conversely that \eqref{eq:correqmainchar} holds for all continuous $f_i^j$.  Fix any measurable $f_i^j$ satisfying the assumptions of the theorem.  Define a signed measure $\pi_i^j$ on $C_i$ by $\pi_i^j(B_i) = \int_{B_i\times C_{-i}}\left[u_i(t_i^j,s_{-i})-u_i(s)\right]\,d\pi$.  Let $\mu_i = \sum_{j=1}^k \lvert \pi_i^j\rvert$ and fix any $\delta > 0$.  Then by the Lemma \ref{lem:generallusin} there exist continuous functions $g_i^j: C_i\rightarrow [0,1]$ which sum to at most unity and equal the $f_i^j$ except on a set of $\mu_i$ measure at most $\delta$.  Therefore
\begin{equation*}
\begin{split}
&\left\lvert \sum_{j=1}^k\int f_i^j(s_i)\left[u_i(t_i^j,s_{-i})-u_i(s)\right]\,d\pi - \sum_{j=1}^k\int g_i^j(s_i)\left[u_i(t_i^j,s_{-i})-u_i(s)\right]\,d\pi \right\rvert \\ & \leq \sum_{j=1}^k\int \lvert f_i^j(s_i) - g_i^j(s_i)\rvert \,d\pi_i^j\leq 2k\delta,
\end{split}
\end{equation*}
so
\begin{equation*}
\sum_{j=1}^k\int f_i^j(s_i)\left[u_i(t_i^j,s_{-i})-u_i(s)\right]\,d\pi \leq \epsilon + 2k\delta.
\end{equation*}
But $\delta$ was arbitrary, so \eqref{eq:correqmainchar} holds for all measurable $f_i^j$.

Finally assume $C_i\subset\Rm^{k_i}$ for some $k_i$.  If \eqref{eq:correqmainchar} holds for all continuous $f_i^j$, then it holds for all squares of polynomials.  Suppose conversely that it holds for all squares of polynomials.  Let $f_i^j$ be any continuous functions satisfying the assumptions of the theorem and $\delta >0$.  Let $p_i^j$ be polynomials squares which approximate the $f_i^j$ within $\delta$ in the sup norm and satisfy the assumptions of the theorem, as provided by Lemma \ref{lem:contpolyapprox}.  Then
\begin{equation*}
\begin{split}
&\left\lvert \sum_{j=1}^k\int f_i^j(s_i)\left[u_i(t_i^j,s_{-i})-u_i(s)\right]\,d\pi - \sum_{j=1}^k\int p_i^j(s_i)\left[u_i(t_i^j,s_{-i})-u_i(s)\right]\,d\pi \right\rvert \\ & \leq \sum_{j=1}^k\int \lvert f_i^j(s_i) - p_i^j(s_i)\rvert \,d\pi_i^j\leq \delta\sum_{j=1}^k \int d\pi_i^j,
\end{split}
\end{equation*}
so
\begin{equation*}
\sum_{j=1}^k\int f_i^j(s_i)\left[u_i(t_i^j,s_{-i})-u_i(s)\right]\,d\pi \leq \epsilon + \delta\sum_{j=1}^k \int d\pi_i^j.
\end{equation*}
But $\delta$ was arbitrary and the integrals on the right are finite, so \eqref{eq:correqmainchar} holds for all continuous $f_i^j$.
\end{proof}

Several simplifications occur when specializing Theorem \ref{thm:correqmainchar} to the $\epsilon = 0$ case, yielding the following characterization.  We will use the polynomial condition of this corollary in Section \ref{subsec:moment} to develop algorithms for computing (approximate) correlated equilibria.  The characteristic function condition is used to compute extreme correlated equilibria of an example game in \cite{sop:sece}.

\begin{corollary}
\label{cor:correqmultiplierchar}
A joint measure $\pi$ is a correlated equilibrium of a continuous game if and only if
\begin{equation}
\label{eq:correqdef3}
\int f_i(s_i)\left[u_i(t_i,s_{-i}) - u_i(s)\right]\,d\pi(s) \leq 0
\end{equation}
for all $i$ and $t_i\in C_i$ as $f_i$ ranges over any of the following sets of functions from $C_i$ to $[0,\infty)$:
\begin{enumerate}
\item Characteristic functions of measurable sets,
\item Measurable simple functions,
\item Bounded measurable functions,
\item Continuous functions,
\item Squares of polynomials (if $C_i\subset \Rm^{k_i}$ for some $k_i$ for all $i$).
\end{enumerate}
\end{corollary}

\begin{proof}
When $\epsilon = 0$ the $k = 1$ case of equation \eqref{eq:correqmainchar} implies the $k > 1$ cases.  Furthermore $\epsilon = 0$ makes \eqref{eq:correqmainchar} homogeneous, so it is unaffected by positive scaling of the $f_i^j$, which allows us to drop the assumption $f_i\leq 1$.
\end{proof}

Theorem \ref{thm:correqmainchar} also has important topological implications for the structure of $\epsilon$-correlated equilibria.  Recall that the weak* topology on the set of probability distributions $\Delta(C)$ over a compact Hausdorff space is the weakest topology which makes $\pi \mapsto \int f \,d\pi$ a continuous functional whenever $f: C\rightarrow\Rm$ is a continuous function.

\begin{corollary}
\label{cor:epscorreqcompact}
The set of $\epsilon$-correlated equilibria of a continuous game is weak* compact.
\end{corollary}

\begin{proof}
By the continuous test function condition in Theorem \ref{thm:correqmainchar}, the set of $\epsilon$-correlated equilibria is defined by conditions of the form $\int f \,d\pi \leq \epsilon$ where $f$ ranges over continuous functions of the form $\sum_{j=1}^k f_i^j(s_i)\left[u_i(t_i^j,s_{-i})-u_i(s)\right]$.  By definition this presents the set of $\epsilon$-correlated equilibria as the intersection of a family of weak* closed sets.  Hence the set of $\epsilon$-correlated equilibria is a closed subset of $\Delta(C)$.  But $\Delta(C)$ is compact by the Banach-Alaoglu theorem \cite{r:fa}, so the set of $\epsilon$-correlated equilibria is compact.
\end{proof}

\begin{corollary}
\label{cor:limitcorreq}
If $\pi^k$ is a sequence of $\epsilon^k$-correlated equilibria and $\epsilon^k\rightarrow 0$, then the sequence $\pi^k$ has a weak* limit point\footnote{It is important to note that here we use the term limit point to refer to a limit point of a sequence, which is slightly different from a limit point of the underlying set of values which appear in the sequence.  The difference is essentially that the singleton set $\{\pi\}$ has no limit points (in the sense of, say, \cite{m:t}), but we would like to say that $\pi$ is a limit point of the constant sequence $\pi,\pi,\pi,\ldots$.  Rigorously, we say that $\pi$ is a limit point of the sequence $\pi^1,\pi^2,\pi^3,\ldots$ if for any neighborhood $U$ of $\pi$, there are infinitely many indices $i$ such that $\pi^i\in U$.  Equivalently (at least in a Hausdorff space) $\pi$ is a limit point of the sequence if and only if $\pi$ appears infinitely often or $\pi$ is a limit point of the underlying set $\{\pi^k\vert k\in\Nm\}$.} and any such limit point is a correlated equilibrium.
\end{corollary}

\begin{proof}
If there is some $\pi$ such that $\pi^k = \pi$ for infinitely many $k$, then $\pi$ is a limit point of the sequence.  Also $\pi$ is an $\epsilon^k$-correlated equilibrium for arbitrarily small $\epsilon^k$, so it is a correlated equilibrium and we are done.  Otherwise, the sequence $\pi^k$ contains infinitely many points.  The space $\Delta(C)$ with the weak* topology is compact by the Banach-Alaoglu theorem \cite{r:fa}, hence any infinite set has a limit point.  Let $\pi\in\Delta(C)$ be a limit point of the sequence $\pi^k$.  For any $\epsilon > 0$ there exists $k_0$ such that for all $k\geq k_0$, $\pi^k$ is an $\epsilon$-correlated equilibrium.  The set $\Delta(C)$ is Hausdorff \cite{r:fa}, so $\pi$ is also a limit point of the set $\{\pi^k\}_{k\geq k_0}$.  Since the set of $\epsilon$-correlated equilibria is compact by Corollary \ref{cor:epscorreqcompact}, the limit point $\pi$ must be an $\epsilon$-correlated equilibrium for all $\epsilon > 0$, i.e.\ a correlated equilibrium.
\end{proof}

Finally, we consider $\epsilon$-correlated equilibria which are supported on some finite subset.  In this case, we obtain another generalization of Proposition \ref{prop:finitecorreqchar} which we will use in the algorithms presented in Section \ref{sec:adaptivedisc}.

\begin{proposition}
\label{prop:sampledepscorreqchar}
A probability measure $\pi\in\Delta(\tilde{C})$, where $\tilde{C} = \Pi_{j\in I} \tilde{C}_j$ is a finite subset of $C$, is an $\epsilon$-correlated equilibrium of a continuous game if and only if there exist $\epsilon_{i,s_i}$ such that
\begin{equation*}
\sum_{s_{-i}\in\tilde{C}_{-i}} \pi(s)\left[u_i(t_i,s_{-i}) - u_i(s)\right] \leq \epsilon_{i,s_i}
\end{equation*}
for all players $i$, all $s_i\in\tilde{C}_i$, and all $t_i\in C_i$, and
\begin{equation*}
\sum_{s_i\in\tilde{C}_i} \epsilon_{i,s_i} \leq \epsilon
\end{equation*}
for all players $i$.
\end{proposition}

\begin{proof}
If we replace $t_i$ with $\zeta_i(s_i)$ in the first inequality then sum over all $s_i\in\tilde{C}_i$ and combine with the second inequality, we get that
\begin{equation}
\label{eq:sampledepscorreq}
\sum_{s\in\tilde{C}} \pi(s)\left[u_i(\zeta_i(s_i),s_{-i}) - u_i(s)\right] \leq \epsilon
\end{equation}
holds for all $i$ and any function $\zeta_i:\tilde{C}_i\rightarrow C_i$.  This is exactly the definition of an $\epsilon$-correlated equilibrium in the case when $\pi$ is supported on the finite set $\tilde{C}$.

Conversely if $\pi$ satisfies \eqref{eq:sampledepscorreq} for all $\zeta_i:\tilde{C}_i\rightarrow C_i$ then let
\begin{equation*}
\epsilon_{i,s_i} = \max_{t_i\in C_i}\sum_{s_{-i}\in\tilde{C}_{-i}} \pi(s)\left[u_i(t_i,s_{-i}) - u_i(s)\right].
\end{equation*}
For each $s_i\in\tilde{C}_i$, let $\zeta_i(s_i)$ be any $t_i\in C_i$ which achieves this maximum; such a $t_i$ exists by compactness and continuity.  Substituting this $\zeta_i$ into \eqref{eq:sampledepscorreq} shows that $\pi$ satisfies the assumptions of the theorem.
\end{proof}

\section{Computing Correlated Equilibria}
\label{sec:comp}

We focus in this section on developing algorithms that can compute approximate correlated equilibria with arbitrary accuracy.  We consider three types of algorithms, which we will illustrate in turn using the example below.  

\savecounter{correqex1}
\begin{example}
\label{ex:correqex1}
Consider the polynomial game with two players, $x$ and $y$, each choosing their strategies from the interval $C_x = C_y = [-1,1]$.  Their utilities are given by
\begin{equation*}
\begin{split}
u_x(x,y) & = 0.596x^2 + 2.072xy - 0.394y^2 + 1.360x
 -1.200y + 0.554 \text{ and}\\
u_y(x,y) & = -0.108x^2 + 1.918xy - 1.044y^2 - 1.232x
 + 0.842y - 1.886.
\end{split}
\end{equation*}
The coefficients have been selected at random.  This example is convenient, because as Figure \ref{fig:g1moment} shows, the game has a unique correlated equilibrium (the players choose $x=y=1$ with probability one).  For the purposes of visualization and comparison, we will project the computed equilibria and approximations thereof into expected utility space, i.e. we will plot pairs $\left(\int u_x\,d\pi,\int u_y\,d\pi\right)$.
\end{example}

\subsection{Static Discretization Methods}
\label{sec:staticdisc}
The static discretization methods we present here are slow in practice and should be taken as a benchmark against which to compare the methods of later sections.  The techniques in this section are general enough to apply to arbitrary continuous games with finitely many players, so we will not restrict our attention to polynomial games here.

The basic idea of static discretization methods is to select some finite subset $\tilde{C}_i \subset C_i$ of strategies for each player and limit his strategy choice to that set.  Restricting the utility functions to the product set $\tilde{C} = \Pi_{i=1}^n \tilde{C}_i$ produces a finite game, called a \textbf{sampled game} or \textbf{sampled version} of the original continuous game.  The simplest computational approach is then to consider the set of correlated equilibria of this sampled game.  This set is defined by the linear inequalities in Proposition \ref{prop:finitecorreqchar} along with the conditions that $\pi$ be a probability measure on $\tilde{C}$.  
The complexity of this approach in practice depends on the number of points in the discretization.

The question is then: what kind of approximation does this technique yield? In general the correlated equilibria of the sampled game may not have any relation to the set of correlated equilibria of the original game.  The sampled game could, for example, be constructed by selecting a single point from each strategy set, in which case the unique probability measure over $\tilde{C}$ is automatically a correlated equilibrium of the sampled game but is a correlated equilibrium of the original game if and only if the points chosen form a pure strategy Nash equilibrium.  Nonetheless, it seems intuitively plausible that if a large number of points were chosen such that any point of $C_i$ were near a point of $\tilde{C}_i$ then the set of correlated equilibria of the finite game would be ``close to'' the set of correlated equilibria of the original game in some sense, despite the fact that each set might contain points not contained in the other.

To make this precise, we will show how to choose a discretization so that the correlated equilibria of the finite game are $\epsilon$-correlated equilibria of the original game.
\begin{proposition}
\label{prop:staticdisc} Consider a continuous game with finitely many players, strategy sets $C_i$, and payoffs $u_i$.  For any $\epsilon > 0$,  there exists a finite open cover $U_i^1,\ldots,U_i^{l_i}$ of $C_i$ such that if $\tilde{C}_i\subseteq C_i$ is a finite set chosen to contain at least one point from each $U_i^l$, then all correlated equilibria of the finite game with strategy spaces $\tilde{C}_i$ and utilities $u_i|_{\tilde{C}}$ will be $\epsilon$-correlated equilibria of the original game.
\end{proposition}
\begin{proof}
Note that the utilities are continuous functions on a compact set, so for any $\epsilon > 0$ we can choose 
a finite open cover $U_i^1,\ldots,U_i^{l_i}$ such that if $s_i$ varies within one of the $U_i^l$ and $s_{-i}\in C_{-i}$ is held fixed, the value of $u_i$ changes by no more than $\epsilon$.  Let $\tilde{C}$ satisfy the stated assumption and let $\pi$ be any correlated equilibrium of the corresponding finite game.  Then by Proposition \ref{prop:finitecorreqchar},
\begin{equation*}
\sum_{s_{-i}\in\tilde{C}_{-i}} \pi(s)\left[u_i(t_i,s_{-i}) - u_i(s)\right] \leq 0
\end{equation*}
for all $i$ and all $s_i,t_i\in\tilde{C}_i$.  Any $t_i\in C_i$ belongs to the same $U_i^l$ as some $\tilde{t}_i\in\tilde{C}_i$, so
\begin{equation*}
\sum_{s_{-i}\in\tilde{C}_{-i}} \pi(s)\left[u_i(t_i,s_{-i}) - u_i(s)\right] \leq \sum_{s_{-i}\in\tilde{C}_{-i}} \pi(s)\left[u_i\left(\tilde{t}_i,s_{-i}\right) - u_i(s) + \epsilon\right] \leq \epsilon\sum_{s_{-i}\in\tilde{C}_{-i}} \pi(s) = \epsilon.
\end{equation*}
Therefore the assumptions of Proposition \ref{prop:sampledepscorreqchar} are satisfied with $\epsilon_{i,s_i} = \epsilon\sum_{s_{-i}\in\tilde{C}_{-i}}\pi(s)$.
\end{proof}

Though our primary goal here is to compute correlated equilibria, not prove existence, it is worth noting that Proposition \ref{prop:staticdisc}, Corollary \ref{cor:limitcorreq}, and the existence of correlated equilibria in finite games \cite{hs:ece} combine to prove the existence of correlated equilibria in continuous games with finitely many players.  Indeed, this is proven in \cite{hs:ece} (along with the extension to an arbitrary set of players) with an argument along similar lines.  One can view much of the contents of the present paper up to this point as expanding on this argument from \cite{hs:ece}.

The proof of Proposition \ref{prop:staticdisc} shows that if the utilities are Lipschitz functions, such as polynomials, then the $U_i^l$ can in fact be chosen to be balls with radius proportional to $\epsilon$.  If the strategy spaces are $C_i = [-1,1]$ as in a polynomial game, then $\tilde{C}_i$ can be chosen to be uniformly spaced within $[-1,1]$.  In this case $\epsilon = O\left(\frac{1}{d}\right)$ where $d=\max_i\left|\tilde{C}_i\right|$.

\begin{figure}
	\centering
		\includegraphics[width=0.7\textwidth]{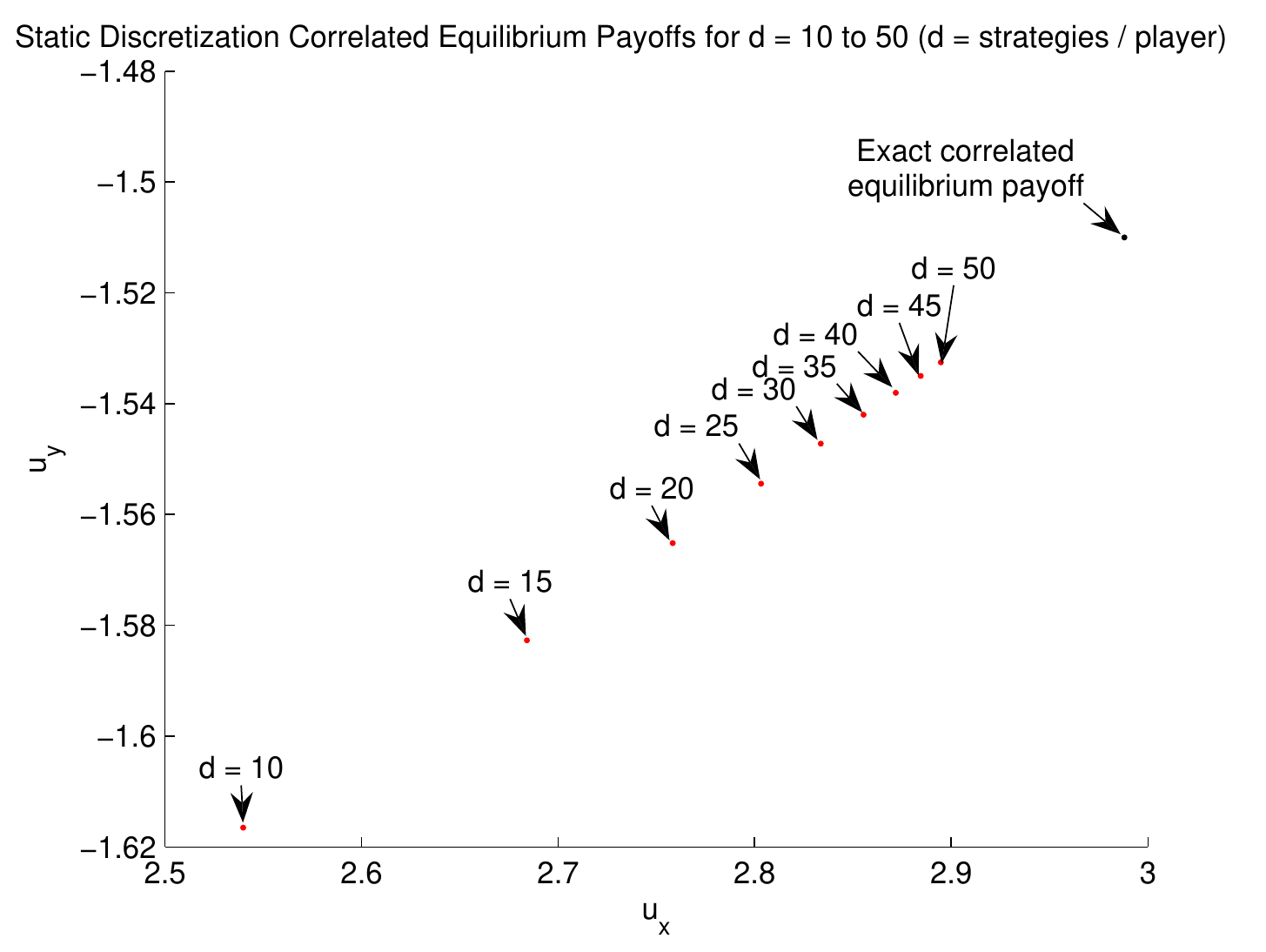}
	\caption{Computing a sequence of $\epsilon$-correlated equilibria of the game in Example \ref{ex:correqex1} by static discretization.  Each point represents the (unique) correlated equilibrium of the finite game where players are restricted to strategies chosen from a finite set of $d$ strategies equally spaced in $[-1,1]$.  The axes represent the utilities received by players $x$ and $y$.  It can be shown that the convergence in this example happens at a rate $\epsilon = \Theta\left(\frac{1}{d}\right)$.  This is slow enough that it is not obvious visually that the sequence of points is converging to the exact correlated equilibrium payoff, though we can prove that it is (e.g., by combining Proposition \ref{prop:staticdisc} with Figure \ref{fig:g1moment}, which shows that the equilibrium payoff is unique).}
		\label{fig:g1static}
\end{figure}

\usesavedcounter{correqex1}
\begin{example}[continued]
Figure \ref{fig:g1static} is a sequence of static discretizations for this game for increasing values of $d$, where $d$ is the number of points in $\tilde{C}_x$ and $\tilde{C}_y$.  These points are selected by dividing $[-1,1]$ into $d$ subintervals of equal length and letting $\tilde{C}_x = \tilde{C}_y$ be the set of midpoints of these subintervals.  For this game it is possible to show that the rate of convergence is in fact $\Theta\left(\frac{1}{d}\right)$ so the worst case bound on convergence rate is achieved in this example.
\end{example}
\restorecounter

In fact we can improve this convergence rate to $\epsilon = O\left(\frac{1}{d^2}\right)$ if we include the endpoints $\pm 1$ in $\tilde{C}_i$ as well and assume that the utilities have bounded second derivatives.  We omit the proof for brevity.

\subsection{Adaptive Discretization Methods}
\label{sec:adaptivedisc}
\subsubsection{A family of convergent adaptive discretization algorithms}
\label{sec:convalgs}
In this section we consider continuous games with finitely many players and provide two algorithms (the second is in fact a parametrized family of algorithms which generalizes the first) to compute a sequence of $\epsilon^k$-correlated equilibria such that $\lim_{k\rightarrow\infty}\epsilon^k = 0$.  By Corollary \ref{cor:limitcorreq} any limit point of this sequence is a correlated equilibrium.  We will show that for polynomial games these algorithms can be implemented efficiently using semidefinite programming.

Informally, these algorithms work as follows.  Each iteration $k$ begins with a finite set $\tilde{C}_i^k\subseteq C_i$ of strategies which each player $i$ is allowed to play with positive probability in that iteration; the initial choice of this set at iteration $k=0$ is arbitrary.  We then compute the ``best'' $\epsilon$-correlated equilibrium in which players are restricted to use only these strategies, i.e., the one which minimizes $\epsilon$ (subject to some extra technical conditions needed to ensure convergence).

Given the optimal objective value $\epsilon^k$ and optimal probability distribution $\pi^k$, there is some player $i$ who can improve his payoff by $\epsilon^k$ if he switches from his recommended strategies to certain other strategies.  We interpret these other strategies as good choices for that player to use to help make $\epsilon^k$ smaller in later iterations $k$.  Therefore we add these strategies to $\tilde{C}_i^k$ to get $\tilde{C}_i^{k+1}$ and repeat this process for iteration $k+1$.

\begin{algorithm}
\label{alg:specificadaptivedisc}
Fix a continuous game with finitely many players.  Let $k = 0$ and for each player fix a finite subset $\tilde{C}_i^0\subseteq C_i$.
\begin{itemize}
\item Let $\pi^k$ be an $\epsilon^k$-correlated equilibrium of the game having minimal $\epsilon^k$ subject to two extra conditions.  First, $\pi^k$ must be supported on $\tilde{C}^k$.  Second, we require that $\pi^k$ be an exact correlated equilibrium of the finite game induced when deviations from the recommended strategies are restricted to the set $\tilde{C}^k$, i.e. when we replace the condition $t_i\in C_i$ in Proposition \ref{prop:sampledepscorreqchar} with $t_i\in\tilde{C}_i^k$.

That is to say, let $\epsilon^k$ be the optimal value of the following optimization problem, and $\pi^k$ be an optimal assignment to the decision variables.
\begin{equation*}
\hskip -0.3in
\begin{array}{rl}
\text{minimize} & \epsilon \\
\text{subject to} & \\
\displaystyle\sum_{s_{-i}\in\tilde{C}_{-i}^k} \pi(s)\left[u_i(t_i,s_{-i}) - u_i(s)\right] \leq 0 & \text{for all } i\text{ and } s_i,t_i\in\tilde{C}_i^k \\
\displaystyle\sum_{s_{-i}\in\tilde{C}_{-i}^k} \pi(s)\left[u_i(t_i,s_{-i}) - u_i(s)\right] \leq \epsilon_{i,s_i} & \text{for all } i\text{, }s_i\in\tilde{C}_i^k \text{and }t_i\in C_i \\
\displaystyle\sum_{s_i\in\tilde{C}_i^k} \epsilon_{i,s_i} \leq \epsilon & \text{for all } i \\
\pi(s) \geq 0 & \text{for all } s\in\tilde{C} \\
\displaystyle\sum_{s\in\tilde{C}^k} \pi(s) = 1 &
\end{array}
\end{equation*}
\item If $\epsilon^k = 0$, terminate.
\item For each player $i$ for whom $\sum_{s_i\in\tilde{C}_i^k} \epsilon_{i,s_i} = \epsilon$, form $\tilde{C}_i^{k+1}$ from $\tilde{C}_i^k$ by adding in,  for each $s_i\in\tilde{C}_i^k$ such that $\epsilon_{i,s_i}>0$, at least one strategy $t_i$ which makes
\[
\sum_{s_{-i}\in\tilde{C}_{-i}^k} \pi(s)\left[u_i(t_i,s_{-i}) - u_i(s)\right] = \epsilon_{i,s_i}.
\]
\item For all other players $i$, let $\tilde{C}_i^{k+1} = \tilde{C}_i^k$.
\item Let $k = k+1$ and repeat.
\end{itemize}
\end{algorithm}

Note that all steps of this algorithm are well-defined.  First, the optimization problem is feasible.  To see this let $\pi^k$ be any exact correlated equilibrium of the finite game with strategy spaces $\tilde{C}_i^k$ and utilities $u_i$ restricted to $\tilde{C}^k$; such an equilibrium exists because all finite games have correlated equilibria \cite{hs:ece}.  The $u_i$ are bounded on $C$ (being continuous functions on a compact set), so by making $\epsilon$ and the $\epsilon_{i,s_i}$ large, we see that $\pi^k$ is a feasible solution of the problem.

Second,  the optimal objective value is achieved by some $\pi^k$ because the space of probability measures on $\tilde{C}^k$ is compact, the constraints are closed, and $\epsilon$ is bounded below by zero.

Third, the set of new strategies added in the third bullet is nonempty.  Suppose for a contradiction that this set were empty for each $i$ such that $\sum_{s_i\in\tilde{C}_i^k} \epsilon_{i,s_i} = \epsilon$ and each $s_i\in\tilde{C}_i^k$ such that $\epsilon_{i,s_i}>0$.  By continuity of $u_i$ and compactness of $C_i$, the left-hand side of the $\epsilon$-correlated equilibrium constraint achieves its maximum as a function of $t_i\in C_i$.  If this maximum value were less than $\epsilon_{i,s_i}$, then the value of $\epsilon_{i,s_i}$ could be decreased.  If this could be done for all $i$ such that $\sum_{s_i\in\tilde{C}_i^k} \epsilon_{i,s_i} = \epsilon$ then $\epsilon$ itself could be decreased, contradicting optimality of $\pi^k$.

Fourth, this set of new strategies added in the third bullet consists only of strategies which are not in $\tilde{C}_i^k$ because we have the constraint that the deviations in utility be nonpositive for $t_i\in\tilde{C}_i^k$.  

To show that Algorithm \ref{alg:specificadaptivedisc} converges, we will view it as a member of the following family of algorithms with the parameters set to $\alpha = 0$ and $\beta = 1$.  Varying these parameters corresponds to adding some slack in the exact correlated equilibrium constraints and allowing some degree of suboptimality in the choice of strategies added to $\tilde{C}_i^k$ to form $\tilde{C}_i^{k+1}$.  Such changes make little conceptual difference, but could be helpful in practice by making the optimization problem strictly feasible and allowing it to be solved to within a known fraction of the optimal objective value rather than all the way to optimality.  We will prove that all algorithms in this family converge, that is, with these algorithms $\epsilon^k$ converges to zero in the limit.

\begin{algorithm}
\label{alg:generaladaptivedisc}
Fix a continuous game with finitely many players and parameters $0\leq \alpha < \beta \leq 1$.  Let $k = 0$ and for each player fix a finite subset $\tilde{C}_i^0\subseteq C_i$.
\begin{itemize}
\item Choose $\epsilon^k$ to be the smallest number for which there exists $\pi^k$ such that:
\begin{itemize}
\item $\pi^k$ is a probability distribution supported on $\tilde{C}^k$,
\item $\pi^k$ is an $\epsilon^k$-correlated equilibrium of the game,
\item $\pi^k$ is not an $\epsilon$-correlated equilibrium for any $\epsilon < \epsilon^k$,
\item $\pi^k$ is an $\alpha\epsilon^k$-correlated equilibrium of the game when strategy deviations are restricted to $\tilde{C}^k$ (i.e., when the condition $t_i\in C_i$ is changed to $t_i\in\tilde{C}_i^k$ in Proposition \ref{prop:sampledepscorreqchar}).
\end{itemize}
\item If $\epsilon^k = 0$, terminate.
\item For at least one value of $i$, form $\tilde{C}_i^{k+1}$ from $\tilde{C}_i^k$ by adding strategies $t_{i,s_i} \in C_i$ such that
\begin{equation*}
\sum_{s\in\tilde{C}^k}\pi^k(s)\left[u_i(t_{i,s_i},s_{-i}) - u_i(s)\right] \geq \beta\epsilon^k.
\end{equation*}
\item For all other values of $i$, let $\tilde{C}_i^{k+1} = \tilde{C}_i^k$.
\item Let $k = k+1$ and repeat.
\end{itemize}
\end{algorithm}

\begin{proposition}
The steps of Algorithm \ref{alg:generaladaptivedisc} are well-defined.
\end{proposition}

\begin{proof}
It is not immediately obvious that the first step of the algorithm is well-defined, i.e., that a minimal $\epsilon^k$ (or any $\epsilon^k$ for that matter) satisfying these conditions exists.  To see this let $\pi^{k,1}$ be an exact correlated equilibrium of the finite game induced when strategy deviations are restricted to $\tilde{C}^k$, and let $\epsilon^{k,1}\geq 0$ be the smallest value such that $\pi^{k,1}$ is an $\epsilon^{k,1}$-correlated equilibrium.  Then the pair $(\pi^{k,1},\epsilon^{k,1})$ satisfies the four conditions under the first bullet above.  This shows that the set of $\epsilon^k$ values satisfying these conditions is nonempty.

Choose some sequence $(\pi^{k,l},\epsilon^{k,l})$, $l = 1,2,\ldots$, of pairs satisfying these conditions such that the limit $\epsilon^k = \lim_{l\rightarrow\infty}\epsilon^{k,l}$ is the infimum over all $\epsilon^k$ values of pairs satisfying these conditions.  Passing to a subsequence if necessary we can assume without loss of generality that the $\pi^{k,l}$ converge to some $\pi^k$.  It is clear from the proof of Corollary \ref{cor:limitcorreq} that $\pi^k$ is an $\epsilon^k$-correlated equilibrium supported on $\tilde{C}^k$ which is an $\alpha\epsilon^k$-correlated equilibrium when deviations are restricted to $\tilde{C}^k$.

From Proposition \ref{prop:sampledepscorreqchar} we see that for a fixed support $\tilde{C}^k$, the minimal value of $\epsilon$ for which a probability measure $\pi$ on $\tilde{C}^k$ is a correlated equilibrium of the game varies continuously with the probabilities $\pi(s)$ for $s\in\tilde{C}^k$.
Therefore $\pi^k$ is not an $\epsilon$-correlated equilibrium for any $\epsilon<\epsilon^k$.  Note that this final step depends crucially on the fact that $\tilde{C}^k$ is finite and fixed while $l$ varies.  Also note that this subtlety disappears if $\alpha = 0$ because in that case it wouldn't matter if the limiting distribution had a smaller $\epsilon$ value.  It is clear that the remaining steps of the algorithm are well-defined.
\end{proof}

\begin{theorem}
\label{thm:adaptconv}Fix a continuous game with finitely many players.  Algorithms \ref{alg:specificadaptivedisc} and \ref{alg:generaladaptivedisc} converge to the set of correlated equilibria, i.e., they converge in the sense that $\epsilon^k\rightarrow 0$.
\end{theorem}

\begin{proof}
Suppose not, so there exists $\epsilon > 0$ and infinitely many values of $k$ such that $\epsilon^k\geq \epsilon$.  For each $i$ let $B_i^1,\ldots, B_i^{l_i}$ be a finite open cover of $C_i$ such that $u_i(s_i,s_{-i}) - u_i(t_i,s_{-i}) \leq \frac{1}{2}(\beta - \alpha)\epsilon$ when $s_i$ and $t_i$ belong to the same set $B_i^l$ and $s_{-i}\in C_{-i}$.  Such a cover exists by the compactness of the $C_i$ and the continuity of the $u_i$.  There are finitely many sets $B_i^l$ so there is some iteration $k$, which we can take to satisfy $\epsilon^k\geq\epsilon$, such that for all $i$ all of the sets $B_i^l$ which will ever contain an element of $\tilde{C}_i^k$ at some iteration $k$ already do.

Note that $\pi^k$ is an $\alpha\epsilon^k$-correlated equilibrium when strategy choices are restricted to $\tilde{C}_i^k$, and $\epsilon^k > 0$ so we have $\beta\epsilon^k>\alpha\epsilon^k$.  By the minimality of $\epsilon^k$, the set $\tilde{C}_i^{k+1}\setminus\tilde{C}_i^k$ is nonempty for some player $i$ (that is to say, it is always possible to perform the third step of the algorithm).  Choose such an $i$ and $t_{i,s_i}\in \tilde{C}_i^{k+1}$ which satisfy
\begin{equation*}
\sum_{s\in\tilde{C}^k} \pi^k(s)\left[u_i(t_{i,s_i},s_{-i}) - u_i(s)\right] \geq \beta\epsilon^k.
\end{equation*}
By assumption, for any choice of $r_{i,s_i}\in\tilde{C}_i^k$ we have
\begin{equation*}
\sum_{s\in\tilde{C}^k} \pi^k(s)\left[u_i(r_{i,s_i},s_{-i}) - u_i(s)\right] \leq \alpha\epsilon^k,
\end{equation*}
so
\begin{equation*}
\sum_{s\in\tilde{C}^k}\pi^k(s)\left[u_i(t_{i,s_i},s_{-i}) - u_i(r_{i,s_i},s_{-i})\right] \geq (\beta - \alpha)\epsilon^k.
\end{equation*}
By construction of $k$, we can choose $r_{i,s_i}\in\tilde{C}_i^k$ to lie in the same set $B_i^l$ as $t_{i,s_i}$ for each $s_i\in\tilde{C}_i^k$.  Thus
\begin{equation*}
\begin{split}
(\beta - \alpha)\epsilon & \leq (\beta - \alpha)\epsilon^k \\ & \leq \left\lvert \sum_{s\in\tilde{C}^k}\pi^k(s)\left[u_i(t_{i,s_i},s_{-i}) - u_i(r_{i,s_i},s_{-i})\right] \right\rvert \\ & \leq \sum_{s\in\tilde{C}^k}\pi^k(s)\left\lvert u_i(t_{i,s_i},s_{-i}) - u_i(r_{i,s_i},s_{-i})\right\rvert \\ &\leq \sum_{s\in\tilde{C}^k} \pi^k(s)\frac{(\beta - \alpha)\epsilon}{2} = \frac{(\beta - \alpha)\epsilon}{2},
\end{split}
\end{equation*}
a contradiction.
\end{proof}

Now we will illustrate Algorithm \ref{alg:specificadaptivedisc} on two examples.

\begin{figure}
	\centering
		\includegraphics[width=.7\textwidth]{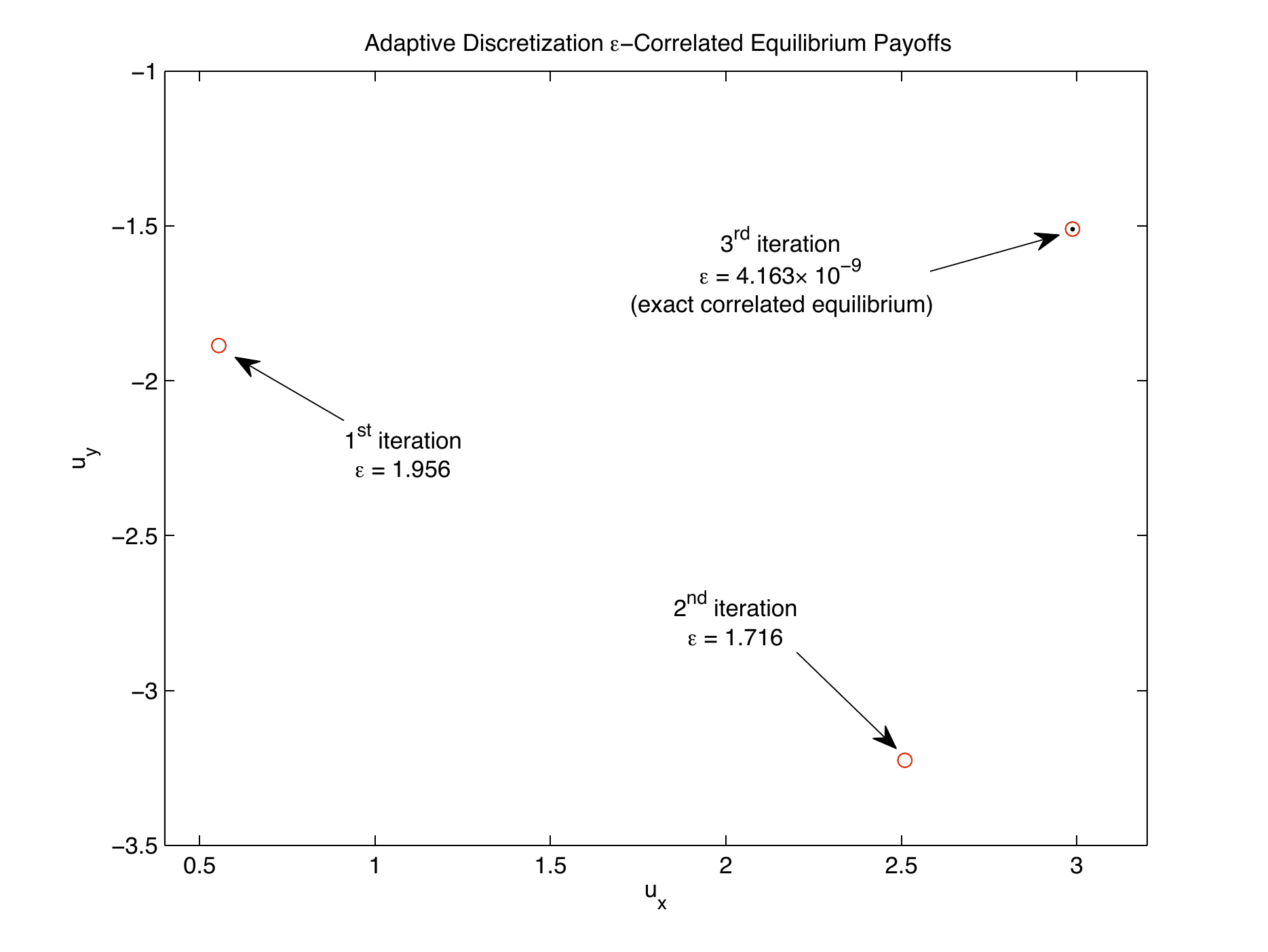}
	\caption{Convergence of Algorithm \ref{alg:specificadaptivedisc} (note the change in scale from Figure \ref{fig:g1static}).  At each iteration, the expected utility pair is plotted along with the computed value of $\epsilon$ for which that iterate is an $\epsilon$-correlated equilibrium of the game.  In this case convergence to $\epsilon = 0$ (to within numerical error) occurred in three iterations.}
	\label{fig:g1adaptive}
\end{figure}

\usesavedcounter{correqex1}
\begin{example}[continued]
In Figure \ref{fig:g1adaptive} we illustrate Algorithm \ref{alg:specificadaptivedisc} initialized with $\tilde{C}_x^0 = \tilde{C}_y^0 = \{0\}$.  In this case convergence is obtained in three iterations, significantly faster than the static discretization method.  The resulting strategy sets were $\tilde{C}_x^2 = \tilde{C}_y^2 = \{0,1\}$.
\end{example}
\restorecounter

\begin{example}
For a more complex illustration, we consider a polynomial game with three players, choosing strategies $x, y,\text{ and }z\in [-1,1]$.  The utilities were chosen to be polynomials with terms up to degree $4$ in all the variables and the coefficients were chosen independently according to a normal distribution with zero mean and unit variance (their actual values are omitted).  Algorithm \ref{alg:specificadaptivedisc} proceeds as in Table \ref{tab:3playerex}, which shows the value of $\epsilon^k$ and the new strategies added on each iteration.  The terminal probability distribution $\pi^6$ does not display any obvious structure; in particular it is not a Nash equilibrium (product distribution).

\begin{table}
\begin{center}
\begin{tabular}{c|c|c|c|c}
$k$ & $\epsilon^k$ & $\tilde{C}_x^k\setminus \tilde{C}_x^{k-1}$ & $\tilde{C}_y^k\setminus\tilde{C}_y^{k-1}$ & $\tilde{C}_z^k\setminus\tilde{C}_z^{k-1}$ \\
\hline
$0$ & $0.99$ & $\{0\} $& $\{0\}$ & $\{0\}$ \\
$1$ & $4.16$ & & & $\{0.89\}$ \\
$2$ & $5.76$ & $\{-1\}$ & & \\
$3$ & $0.57$ & & $\{1\}$ & \\
$4$ & $0.28$ & $\{0.53\}$ & & $\{0.50,0.63\}$ \\
$5$ & $0.16$ & & $\{0.49,0.70\}$ & \\
$6$ & $10^{-7}$& & $\{-1,0.60\}$ & $\{-0.60,0.47\}$ \\
\end{tabular}
\end{center}
\caption{Output of Algorithm \ref{alg:specificadaptivedisc} on a three player polynomial game with utilities of degree $4$ and randomly chosen coefficients.}
\label{tab:3playerex}
\end{table}
\end{example}

\subsubsection{Implementing these algorithms with semidefinite programs}
\label{sec:imp}
To implement these algorithms for polynomial games, we must be able to do two things.  First, we need to solve optimization problems with finitely many decision variables, linear objective functions and two types of constraints: nonnegativity constraints on linear functionals of the decision variables, and nonnegativity constraints on univariate polynomials whose coefficients are linear functionals of the decision variables.  That is to say, we must be able to handle constraints of the form $p(t)\geq 0$ for all $t\in [-1,1]$, where the coefficients of the polynomial $p$ are linear in the decision variables.  Second, we need to extract values of $t$ for which such polynomial inequalities are tight at the optimum.

Both of these tasks can be done simultaneously by casting the problem as a \textbf{semidefinite program (SDP)}.  For an overview of semidefinite programs and a summary of the necessary results (both of which are classical), see the appendix.

In the optimization problem in Algorithm \ref{alg:specificadaptivedisc} we have a finite number of univariate polynomials in $t_i$ whose coefficients are linear in the decision variables $\pi(s)$ and $\epsilon_{i,s_i}$.  We wish to constrain these coefficients to allow only polynomials which are nonnegative for all $t_i\in [-1,1]$.  By Propositions \ref{prop:posintsos} and \ref{prop:sossdp} in the appendix this is the same as asking that these coefficients equal certain linear functions of matrices (i.e., sums along antidiagonals) which are constrained to be symmetric and positive semidefinite.  Therefore we can write this optimization problem as a semidefinite program.

As a special case of convex programs, semidefinite programs have a rich duality theory which is useful for theoretical and computational purposes.  In particular, SDP solvers keep track of feasible primal and dual solutions in order to determine when optimality is reached.  It can be shown that the dual data obtained by an SDP solver run on this optimization problem will encode the values of $t_i$ making the polynomial inequalities tight at the optimum \cite{p:phd}.

The process of generating an SDP from the optimization problem in the algorithms above, solving it, and extracting an optimal solution along with $t_i$ values from the dual can all be automated.  We have done so using the SOSTOOLS MATLAB toolbox for the pre- and post-processing and SeDuMi for solving the semidefinite programs efficiently \cite{sostools, sedumi}.

\subsubsection{A nonconvergent limiting case}
\label{sec:nonconvalgs}
Note that in the algorithms above the convergence of the sequence $\epsilon^k$ is not necessarily monotone.  If we were to let $\alpha = \beta$ (a case we did not allow above), the sequence would become monotone nonincreasing.  If we were to furthermore fix $\alpha = \beta = 1$, then the condition that $\pi$ be an exact (or $\alpha\epsilon^k$-) correlated equilibrium when deviations are restricted to $\tilde{C}_i^k$ would become redundant and could be removed.

These changes would simplify the behavior of Algorithm \ref{alg:generaladaptivedisc} conceptually as well as reducing the size of the SDP solved at each iteration, so we would like to adopt them if possible.  However, the resulting algorithm may not converge, in the sense that $\epsilon^k$ may remain bounded away from zero.

\begin{table}
\begin{center}
\begin{tabular}{c|c|c|c|}
& $a$ & $b$ & $c$ \\
\hline 
$a$ & $0$ & $1$ & $0$ \\
\hline
$b$ & $1$ & $5$ & $7$ \\
\hline
$c$ & $0$ & $7$ & $0$ \\
\hline
\end{tabular}
\caption{A finite symmetric game with identical utilities for which Algorithm \ref{alg:generaladaptivedisc} with $\alpha = \beta = 1$ does not converge when started with strategy sets $\tilde{C}_1^0 = \tilde{C}_2^0 = \{a\}$.}
\label{tab:adaptivediscbad}
\end{center}
\end{table}

\begin{example}
Consider the game shown in Table \ref{tab:adaptivediscbad}, which is symmetric and has identical utilities for both players.  Let $\tilde{C}_1^0 = \tilde{C}_2^0 = \{a\}$ and apply Algorithm \ref{alg:specificadaptivedisc}, but remove the condition that $\pi^k$ be an exact correlated equilibrium when deviations are restricted to $\tilde{C}_i^k$.  The only probability distribution supported on $\tilde{C}^0$ is $\delta_{(a,a)}$ which has an objective value of $\epsilon^0 = 1$.  It is easy to see that $\tilde{C}_i^1$ is formed by simply adding each player's best response to $a$, so that $\tilde{C}_1^1 = \tilde{C}_2^1 = \{a,b\}$.  We will argue that the unique solution to the optimization problem in iteration $k = 1$ is also $\delta_{(a,a)}$, hence $\tilde{C}_i^2 = \tilde{C}_i^1$ and the algorithm gets ``stuck'', so that $\epsilon^k = \epsilon^0 = 1$ for all $k$.

For a probability distribution $\pi$, let $\pi^{T}$ denote $\pi$ with the players interchanged.  By symmetry and convexity, if $\pi$ is an optimal solution then so is $\frac{\pi+\pi^T}{2}$, which is a symmetric probability distribution with respect to the two players.  Hence an optimal solution which is symmetric always exists.  We will parametrize such distributions by $\pi = p\delta_{(a,a)} + q\delta_{(a,b)} + q\delta_{(b,a)} + r\delta_{(b,b)}$, where $p,q,r\geq 0$ and $p + 2q + r = 1$.  Define a departure function $\zeta: C_1\rightarrow C_1$ by $\zeta(a) = b$, $\zeta(b) = \zeta(c) = c$.  Then for $\pi$ to be an $\epsilon$-correlated equilibrium it must satisfy the following condition:
\begin{equation*}
\begin{split}
\epsilon & \geq \sum_{s_1\in \tilde{C}_1^1} \epsilon_{1,s_1} \geq \sum_{s \in \tilde{C}^1} \pi(s)\left[u_1(\zeta(s_1),s_2) - u_1(s_1,s_2)\right] \\
& = p + 4q - q + 2r = 1 + q + r.
\end{split}
\end{equation*}
We know we can achieve $\epsilon = 1$ with $p = 1$ (i.e. $\pi = \pi^0 = \delta_{(a,a)}$), and this inequality shows that if $p < 1$ then $\epsilon > 1$.  Therefore the minimal $\epsilon$ value in iteration $k = 1$ is unity and is achieved by $\pi = \delta_{(a,a)}$.  Furthermore we have shown that this is the unique symmetric probability distribution which achieves the minimal value of $\epsilon$.  Hence any other (not necessarily symmetric) optimal solution $\hat{\pi}$ satisfies $\frac{\hat{\pi} + \hat{\pi}^T}{2} = \delta_{(a,a)}$.  But $\delta_{(a,a)}$ is an extreme point of the convex set of probability distributions on $\tilde{C}^1$, so we must in fact have $\hat{\pi} = \delta_{(a,a)}$.  Therefore $\pi^1 = \pi^0 = \delta_{(a,a)}$ is the unique optimal solution on iteration $k = 1$, so the procedure must get stuck as claimed.  That is, $\tilde{C}_i^k = \{a,b\}$ and $\epsilon^k = 1$ for all $k\geq 1$. 
\end{example}

The same behavior can occur in polynomial games, as can be shown by ``embedding'' the above finite game in a polynomial game.  For example, we can take $C_x = C_y = [-1,1]$ and
\begin{equation*}
\begin{split}
u_x(x,y) = u_y(x,y) =\ & (1-x^2)(3y^2 + 6y + 5) \\ +\ & (1-y^2)(3x^2+6x+5).
\end{split}
\end{equation*}
Then if $\tilde{C}_x^0 = \tilde{C}_y^0 = \{-1\}$ the same analysis as above shows that $\tilde{C}_x^k = \tilde{C}_y^k = \{-1,0\}$ and $\epsilon^k = 2$ for all $k\geq 1$.

\begin{example}
If we run Algorithm \ref{alg:specificadaptivedisc} on this polynomial game, the iterations proceed as in Table \ref{tab:2playerex}.  The correlated equilibrium obtained in iteration $2$ is
\begin{equation*}
\begin{split}
\pi^2 = &\ 0.4922\delta(x=0,y=1) + 0.4922\delta(x=1,y=0) \\
&+ 0.0156\delta(x=1,y=1),
\end{split}
\end{equation*}
i.e., a probability of $0.4922$ is assigned to each of the outcomes $(x,y) = (0,1)$ and $(x,y) = (1,0)$ and a probability of $0.0156$ is assigned to $(x,y) = (1,1)$.

\begin{table}
\begin{center}
\begin{tabular}{c|c|c|c}
$k$ & $\epsilon^k$ & $\tilde{C}_x^k\setminus \tilde{C}_x^{k-1}$ & $\tilde{C}_y^k\setminus\tilde{C}_y^{k-1}$ \\
\hline
$0$ & $2$ & $\{-1\}$ & $\{-1\}$ \\
$1$ & $4$ & $\{0\}$ & $\{0\}$ \\
$2$ & $0$ & $\{1\}$ & $\{1\}$ \\
\end{tabular}
\end{center}
\caption{Output of Algorithm \ref{alg:specificadaptivedisc} for a polynomial game on which Algorithm \ref{alg:generaladaptivedisc} with $\alpha = \beta = 1$ does not converge to a correlated equilibrium.}
\label{tab:2playerex}
\end{table}
\end{example}

\subsection{Moment Relaxation Methods}
\label{subsec:moment}
In this section we again consider only polynomial games.  The \textbf{moment relaxation methods} for computing correlated equilibria have a different flavor from the discretization methods discussed above.  Instead of using tractable finite approximations of the correlated equilibrium problem derived via discretizations, we begin with the alternative exact characterization given in condition \ref{item:poly} of Corollary \ref{cor:correqmultiplierchar}.  In particular, a measure $\pi$ on $C$ is a correlated equilibrium if and only if
\begin{equation}
\label{eq:correqpolychar}
\int p^2(s_i)\left[u_i(t_i,s_{-i}) - u_i(s)\right]\,d\pi(s) \leq 0
\end{equation}
for all $i$, $t_i\in C_i$, and polynomials $p$.  If we wish to check all these conditions for polynomials $p$ of degree less than or equal to $d$, we can form the matrices
\begin{equation*}
S_i^d = \left[\begin{array}{ccccc}1 & s_i & s_i^2 & \cdots & s_i^d \\s_i & s_i^2 & s_i^3 & \cdots & s_i^{d+1} \\s_i^2 & s_i^3 & s_i^4 & \cdots & s_i^{d+2} \\\vdots & \vdots & \vdots & \ddots & \vdots \\s_i^d & s_i^{d+1} & s_i^{d+2} & \cdots & s_i^{2d}\end{array}\right].
\end{equation*}
Let $c$ be a column vector of length $d+1$ whose entries are the coefficients of $p$, so $p^2(s_i) = c' S_i^d c$.  If we define
\begin{equation*}
M_i^d(t_i) = \int S_i^d\left[u_i(t_i,s_{-i}) - u_i(s)\right] \,d\pi(s),
\end{equation*}
then \eqref{eq:correqpolychar} is satisfied for all $p$ of degree at most $d$ if and only if $c' M_i^d(t_i) c\leq 0$ for all $c\in\Rm^{d+1}$ and $t_i\in C_i$, i.e. if and only if $M_i^d(t_i)$ is negative semidefinite for all $t_i\in C_i$.

The matrix $M_i^d(t_i)$ has entries which are polynomials in $t_i$ with coefficients which are linear in the joint moments of $\pi$.  By Proposition \ref{prop:psdinterval} in the appendix, $M_i^d(t_i)$ is negative semidefinite for all $t_i\in [-1,1]$ for a given $d$ and a fixed $\pi$ if and only if there exists a certificate of a certain form proving this condition holds.  We can write a semidefinite program (again, see Proposition \ref{prop:psdinterval} in the appendix) in which the decision variables represent such a certificate, so we can check this condition by solving the semidefinite program.  As $d$ increases we obtain a sequence of semidefinite relaxations of the correlated equilibrium problem and these converge to the exact condition for a correlated equilibrium.  That is to say, for a measure to be a correlated equilibrium it is necessary and sufficient that its moments be feasible for all of these semidefinite programs.

We can also let the measure $\pi$ vary by replacing the moments of $\pi$ with variables and constraining these variables to satisfy some necessary conditions for the moments of a joint measure on $C$ (see appendix).  These conditions can be expressed in terms of semidefinite constraints and there is a sequence of these conditions which converges to a description of the exact set of moments of a joint measure $\pi$.  Thus we obtain a nested sequence of semidefinite relaxations of the set of moments of measures which are correlated equilibria, and this sequence converges to the set of correlated equilibria.

\begin{figure}
	\centering
		\includegraphics[width=0.7\textwidth]{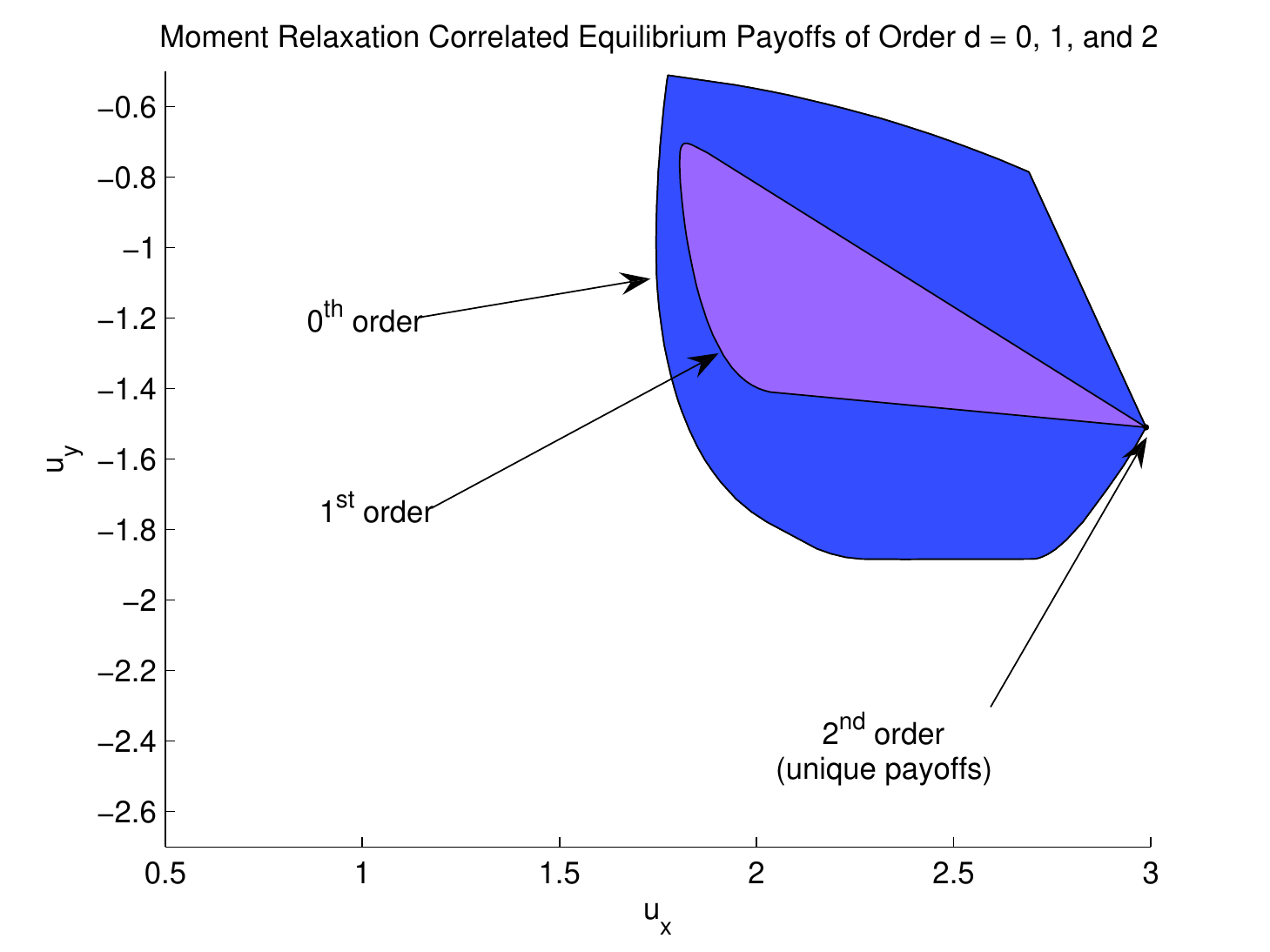}
	\caption{Semidefinite relaxations approximating the set of correlated equilibrium payoffs.  The second order relaxation is a singleton, so this game has a unique correlated equilibrium payoff (and in fact a unique correlated equilibrium).}
	\label{fig:g1moment}
\end{figure}

\usesavedcounter{correqex1}
\begin{example}[continued]
Figure \ref{fig:g1moment} shows moment relaxations of orders $d=0,1,\text{ and }2$.  Since moment relaxations are outer approximations to the set of correlated equilibria (having been defined by necessary conditions which correlated equilibria must satisfy) and the $2^{\text{nd}}$ order moment relaxation corresponds to a unique point in expected utility space, all correlated equilibria of the example game have exactly this expected utility.  In fact, the set of points in this relaxation is a singleton (even before being projected into utility space), so this proves that this example game has a unique correlated equilibrium.
\end{example}
\restorecounter

\section{Future Work}
These results leave several open questions.  For any continuous game, the set of correlated equilibria is nonempty, and this can be proven constructively as in \cite{hs:ece}.  Under the same assumptions we can prove the existence of a Nash equilibrium, but the proof is nonconstructive, or at least does not seem to give an efficient algorithm for constructing an equilibrium \cite{sop:slrcg}.  In the case of polynomial games, existence of a Nash equilibrium immediately gives existence of a finitely supported Nash equilibrium by Carath\'{e}odory's theorem, which is constructive \cite{sop:slrcg}.  Therefore there exists a finitely supported correlated equilibrium of any polynomial game.  Is there a constructive way to prove this fact directly, without going through Nash equilibria?  Such a proof could potentially lead to a provably efficient algorithm for computing a sample correlated equilibrium of a polynomial game.

While the adaptive discretization and moment relaxation algorithms converge in general and work well in practice, we do not know of any results regarding rate of convergence.  If we regard the probability distributions produced by these algorithms at the $k^{\text{th}}$ iteration as $\epsilon^k$-correlated equilibria, how fast does $\epsilon^k$ converge to zero?

Finally, we note that we have merely shown that the adaptive discretization algorithm converges to the set of correlated equilibria, not to a particular correlated equilibrium (of course it will do so along some subsequence by compactness).  Could the algorithm be modified to converge to a single correlated equilbrium?  Or even better, could one assure convergence to a correlated equilibrium with some desirable properties, such as one which maximizes the social welfare or (in the polynomial case) is finitely supported?  This seems plausible given that the algorithm is itself optimization-based, but these problems remain open.

\section*{Acknowledgements}
The authors would like to thank Professor Muhamet Yildiz for a productive discussion which led to an early formulation of the characterization theorems in Section \ref{sec:char} as well as the moment relaxation methods presented in Section \ref{subsec:moment}.  Figures were produced using the SeDuMi package for MATLAB \cite{sedumi}.

\appendix
\section{Semidefinite programming, sums of squares, and moments of measures}
\begin{definition}A \textbf{semidefinite program} is an optimization problem of the form:

\begin{center}
\begin{tabular}{rl}
minimize & $L(S)$ \\
subject to & $T(S) = v$ \\
& $S$ is a symmetric matrix \\
& $S\succeq 0$ (positive semidefinite),
\end{tabular}
\end{center}
where $L$ is a given linear functional, $T$ is a given linear transformation, $v$ is a given vector, and $S$ is a square matrix of decision variables.
\end{definition}

Semidefinite programs are convex optimization problems and generalize linear programs ($T$ and $v$ can be designed to make $S$ diagonal, in which case the condition $S\succeq 0$ is the same as the condition that $S\geq 0$ elementwise).  The solution set of a semidefinite program need not be polyhedral, allowing for much more flexibility in modeling than can be achieved with linear programs.  Many problems can be expressed exactly or approximately as semidefinite programs, and this is important because semidefinite programs can be solved efficiently by interior point methods.  For details and a variety of examples see \cite{vb:sdp} and \cite{p:phd}.


The square of a real-valued function is nonnegative on its entire domain, as is a sum of squares of real-valued functions.  In particular, any polynomial of the form $p(x) = \sum p_k^2(x)$, where $p_k$ are polynomials, is guaranteed to be nonnegative for all $x$.  This gives a sufficient condition for a polynomial to be nonnegative.  It is a classical result that this condition is also necessary if $p$ is univariate \cite{r:scah17p}.

\begin{proposition}
\label{prop:unisos}
A univariate polynomial $p$ is nonnegative on $\Rm$ if and only if it is a sum of squares.
\end{proposition}
\begin{proof}A simpler version of the proof of the following proposition.
\end{proof}

\begin{proposition}[Markov-Luk\'{a}cs \cite{kn:mmpep}]
\label{prop:posintsos} A univariate polynomial $p(x)$ is nonnegative on the interval $[-1,1]$ if and only if $p(x) = s(x) + (1-x^2)t(x)$ where $s$ and $t$ are both sums of squares of polynomials.
\end{proposition}

\begin{proof}
Direct algebraic manipulations show that the set of polynomials of the form $s(x) + (1-x^2)t(x)$ where $s$ and $t$ are sums of squares of polynomials in $x$ is closed under multiplication and contains all polynomials of the following forms: $a$ for $a\geq 0$, $(x-a)^2 + b^2$ for $a,b\in\Rm$, $x - a$ for $a \leq -1$, and $a - x$ for $a \geq 1$.  By assumption $p(x)$ factors as a product of terms of these types, because any real root of $p$ in the interval $(-1,1)$ must have even multiplicity.
\end{proof}

These sum of squares conditions are easy to express using linear equations and semidefinite constraints.  

\begin{proposition}
\label{prop:sossdp}
A univariate polynomial $p(x) = \sum_{k=0}^{2d} p_k x^k$ of degree at most $2d$ is a sum of squares of polynomials if and only if there exists a symmetric positive semidefinite matrix $Q\in\Rm^{(d+1)\times(d+1)}$ such that $p_k = \sum_{i+j=k} Q_{ij}$ (numbering the rows and columns of $Q$ from $0$ to $d$).
\end{proposition}
\begin{proof}
Relating the coefficients of $p(x)$ to the entries of $Q$ in this way is the same as writing $p(x) = \mathbf{x}^TQ\mathbf{x}$ where $\mathbf{x} = \begin{bmatrix}1 & x & x^2 & \cdots & x^d\end{bmatrix}^T$.  Thought of in this way, saying that $p(x)$ is a sum of squares is the same as saying that $Q = \sum_i q_iq_i^T$ for some column vectors $q_i$ and in this case $Q$ is clearly positive semidefinite.  Conversely, if $Q$ is positive semidefinite then there exists a matrix $F$ such that $Q = F^TF$, so $p(x) = \mathbf{x}^TQ\mathbf{x} = \sum_i \left[F\mathbf{x}\right]_i^2$.
\end{proof}

Similar semidefinite characterizations exist for multivariate polynomials to be sums of squares.  While the condition of being a sum of squares does not characterize general nonnegative multivariate polynomials exactly, there exist sequences of sum of squares relaxations which can approximate the set of nonnegative polynomials (on e.g. $\Rm^k$, $[-1,1]^k$, or a more general semialgebraic set) arbitrarily tightly \cite{r:scah17p}.  Furthermore, for some special classes of multivariate polynomials, the sum of squares condition is exact.  

\begin{proposition}
\label{prop:psdinterval}
A matrix $M(t)$ whose entries are univariate polynomials in $t$ is positive semidefinite on $[-1,1]$ if and only if $x'M(t)x = S(x,t) + (1-t^2)T(x,t)$ where $S$ and $T$ are polynomials which are sums of squares.
\end{proposition}

\begin{proof} Follows from Theorem $5.6$ of \cite{clr:rzpsdf}.
\end{proof}

Now suppose we wish to answer the question of whether a finite sequence $(\mu^0,\ldots,\mu^k)$ of reals correspond to the moments of a measure on $[-1,1]$, i.e. whether there exists a positive measure $\mu$ on $[-1,1]$ such that $\mu^i = \int x^i \,d\mu(x)$.  Clearly if such a measure exists then we must have $\int p(x)\,d\mu(x)\geq 0$ for any polynomial $p$  of degree at most $k$ which is nonnegative on $[-1,1]$.  
This necessary condition for moments to correspond to a measure turns out to be sufficient \cite{ks:gms} and can be written in terms of semidefinite constraints.

\begin{proposition}
The condition that a finite sequence of numbers $(\mu^0,\ldots,\mu^k)$ be the moments of a positive measure on $[-1,1]$ can be written in terms of linear equations and semidefinite matrix constraints.
\end{proposition}

One can formulate similar questions about whether a finite sequence of numbers corresponds to the joint moments $\int x_1^{i_1}\cdots x_k^{i_k}\,d\mu(x)$ of a positive measure $\mu$ on $[-1,1]^k$ (or a more general semialgebraic set).  Using a sequence of semidefinite relaxations of the set of nonnegative polynomials on $[-1,1]^k$, a sequence of necessary conditions for joint moments is obtained.  
These conditions approximate the set of joint moments arbitrarily closely.

\bibliographystyle{plain}
\bibliography{../../references}
\end{document}